\RequirePackage[2020-02-02]{latexrelease}
\documentclass[amssymb,twocolumn,aps]{revtex4}

\draft 
\usepackage{natbib}
\usepackage{graphicx}
\usepackage{amsmath}
\usepackage{anyfontsize} 
\usepackage{float}
\usepackage[caption = false]{subfig}
\usepackage{url}
\usepackage{dcolumn}
\usepackage{bm}
\usepackage[utf8]{inputenc}
\usepackage[T1]{fontenc}
\usepackage{mathptmx}
\usepackage{etoolbox}
\begin{document}


\title{Study of Low-dimensional Nonlinear 
	Fractional Difference Equations of Complex Order} 



\author{Divya D Joshi}
\email[]{divyajoshidj27@gmail.com}
\affiliation{Department of Physics, Rashtrasant Tukadoji Maharaj Nagpur University, Nagpur, India- 440033}

\author{Prashant M Gade}
\email[corresponding author: ]{prashant.m.gade@gmail.com}
\affiliation{Department of Physics, Rashtrasant Tukadoji Maharaj Nagpur University, Nagpur, India- 440033}

\author{Sachin Bhalekar}
\email[]{sachinbhalekar@uohyd.ac.in}
\affiliation{School of Mathematics and Statics, University of Hyderabad, Hyderabad, India- 500046}


\date{\today}

\begin{abstract}
We study the fractional maps of complex 
order,
$\alpha_0e^{i r \pi/2}$
for $0<\alpha_0<1$ and $0\le r<1$  in 1 and 2 dimensions. 
In two dimensions, we study H{\'e}non and Lozi map and in $1d$, we
study 
logistic, tent, Gauss, circle, and Bernoulli maps.
The generalization
in $2d$ can be done in two
different ways which are not equivalent for fractional-order
and lead to different bifurcation diagrams.
We observed that the smooth maps such as logistic, Gauss, and H{\'e}non
maps do not show chaos while discontinuous maps such as
Lozi, Bernoulli, and circle maps show chaos. The tent map is continuous
but not differentiable and it shows chaos as well.
In $2d$, we find that the complex fractional-order
maps that show chaos also show multistability.
Thus, it can be inferred that the smooth maps of
complex fractional-order tend to show more regular behavior
than the discontinuous or non-differentiable  maps.

\end{abstract}


\maketitle 

\begin{quotation}
\textbf{Complex differential equations of real fractional order or
	 differential equations of complex fractional order have been studied 
	 in the context of some applications. We study the dynamics of 
	 difference equations of complex fractional order. In general, 
	 the right-hand side of these equations involves arbitrary functions. 
	 We find that for a highly restrictive set of functions, {\it{i.e.}}, 
	 complex analytic functions, no chaos is observed in dynamics. The 
	 variables can be extended to complex space by using complex initial 
	 conditions even for real fractional order. We observe that complex 
	 difference equations of real fractional order do not show any chaos 
	 for complex analytic functions either.}
\end{quotation}

\section{Introduction}
\noindent

Though studies in fractional calculus started from Leibniz and 
almost all leading mathematicians have contributed
to its theory, it has received 
tremendous attention in the last few decades.
Fractional versions
of several differential equations have been investigated
numerically as well as analytically.
These studies are mostly related to real fractional-order
differential equations. 
It has found several applications in the recent past in fields as diverse
as heat transfer equations and viscoelasticity and is an active area of
research \cite{francisco2014fractional,oprzkedkiewicz2021fractional,meral2010fractional}. 
A natural curiosity is whether we can obtain a derivative of
imaginary order and Love can be credited for defining it for the
first time \cite{love1971fractional}. This was later extended to
arbitrary complex
order \cite{andriambololona2012definitions,campos1990solution}.
Unlike real fractional order differential equations, applications of complex fractional order
differential equations are not so well established. Makris and Constantinou
suggested applications in viscoelasticity \cite{makris}. Makris also gave
a complex parameter Kevin model for  elastic foundations \cite{makris2}.
The boundary value problem for fractional differential equation
of complex order is studied in \cite{nea}.
As noted in \cite{ortigueira2021complex}, a
 major difficulty with
complex derivatives is that it treats positive and negative frequencies
differently and the sum or difference of complex order derivative and
its conjugate has been proposed as one of the solutions \cite{adams,atanackovic2015vibrations}.
Still, the complex order fractional derivatives can be 
quite attractive from the application point of view. As
stated by Makris, "Complex-parameter models are very
attractive, because a minimum number of parameters is required to obtain a satisfactory fit of the ‘exact’
response. For instance, in modeling the response of a rigid disc resting on elastic foundations, only two
complex-valued frequency-independent parameters are sufficient to reproduce closely the rigorously 
obtained dynamic stiffness for the vertical, horizontal, and rocking modes" \cite{makris3}.
Atanovic and Pilipovic studied heat conduction with a
general form of 
a constitutive equation containing a
fractional derivative of real and complex
order
\cite{atanackovic2018constitutive}.

From the viewpoint of control, fractional order controllers 
are found to be
effective and complex order controllers have also been used. It is obvious that
complex order operators produce complex-valued output even for a real-valued
function and hence it was proposed that a complex-order operator should be
paired with conjugate order operator\cite{Hartley}.
Tare et al. designed complex-order PID controller structures 
using conjugated order derivatives. The model 
was comparatively studied and found to have an overall better performance owing to its more flexible 
structure \cite{tare2019design}. Sekhar {\it{et al}} studied complex order 
controller where they simply omitted the imaginary part of derivative as well as complex integration\cite{sekhar2020complex}.
Another interesting application of complex order derivative has been 
particle swarm optimization\cite{pahnehkolaei2021particle}.

Nonlinear dynamics of continuous-time complex order
fractional systems has also received some attention. Pinto has
studied the dynamics of two  unidirectional rings of cells coupled through
a buffer cell\cite{pinto2015strange}. Applications of complex order derivative 
have been studied in animal locomotion\cite{pinto2011complex}. The theory 
of hybrid fractional differential equations with complex order
has been developed in \cite{vivek2019theory}.
In \cite{pinto}, Pinto and Carvalho studied the fractional 
complex order model for HIV infection. Variation of complex order 
sheds light on the modeling of intracellular delay and the 
model offers rich dynamics.
Pinto and Machado studied complex order van der Pol oscillator 
and complex order
forced van der Pol oscillator \cite{pinto2, pinto3}.

The 
introduction of a complex-order fractional derivative leads to complex-valued
variables even if initial conditions are real. We may also 
study the dynamics of complex fractional differential equations 
where the real fractional-order derivative is employed. There are several
works in this context. Chaos synchronization, as well as control, 
has been observed
 in fractional-order complex, chaotic systems, and it has been studied in
  \cite{singh2017synchronization,luo2013chaos,gao2005chaos}.
In all these works, we observe that the functions are functions of the variables 
as well as their conjugates. Thus, these are not analytic functions and they are 
effectively real dynamical systems with double the dimension. It would
be interesting to study the possibility of chaos if we insist that the functions
should be analytic. We explore this question
in the context of discrete maps by studying several systems.

Discrete maps have played
a major role in understanding dynamical systems. Simulations of discrete maps
are easier computationally. Several phenomena that appear in flows
occur in maps as well \cite{ott}. Thus, understanding discrete maps 
can complement our understanding of flows.
Many control schemes useful for
the control of differential equations can be used for maps as 
well \cite{shinbrot1993using}.

While we need at least a three-dimensional continuous-time system to
observe chaos, it can be observed even in one-dimensional difference
equations. The most-studied maps in this context are logistic
and tent maps. The question is whether this feature is retained in presence 
of memory and studies in fractional difference equations are important
in this context.
The studies in fractional difference equations are relatively
recent \cite{holm2011theory,atici2009initial,atici2007transform,atici2012gronwall}. 
There have been studies on the stability of fractional difference equations
and in recent times, the definition is extended to
complex orders \cite{gade2021fractional,bhalekar2022stability}. In particular,
the stability conditions for linear fractional
difference equations of
complex order have been derived \cite{bhalekar2022stability}. 
For this scheme, linear systems have been studied in \cite{gade2021fractional}
and it is known that the trajectories that converge to origin
converge as a Mittag-Leffler function which is a power-law
asymptotically. On the other hand, unstable trajectories 
diverge exponentially. Thus, positive Lyapunov exponent
can be used to quantify chaos in these systems.
It is of interest to explore the possibility of
chaos in genuinely complex maps in any dimension, where the functions 
are analytic functions of variables and do
not involve their complex conjugates.

We briefly review prototypical and well-studied systems in 
discrete dynamics.
The logistic map and tent map in $1d$ show 
chaos and have
been extensively studied in this context. 
The maps, such as the 
circle map, 
model dynamical systems represented by the damped driven pendulum such as
Josephson junction in the microwave field, 
charge density waves, lasers, and even 
air-bubble formation 
\cite{bohr1984transition,tredicce1985instabilities,detienne1997semiconductor,tufaile2001circle}. Gauss map has domain over $\mathbb{R}$,  unlike other maps.
The Bernoulli map is easy to study analytically. These maps have been
studied by several researchers. In $2d$, the H{\'e}non map and Lozi map are 
popular and well-investigated maps.
We study the extension of these $1d$ and $2d$ maps to 
fractional complex orders.
These systems can be classified into a few different 
categories. Logistic map, H{\'e}non map, and Gauss map are
continuous and differentiable. Bernoulli, circle,  and Lozi maps are 
discontinuous. The tent map is continuous but not differentiable. 
The key finding is that continuous and differentiable maps do
not show chaotic attractors for complex fractional orders. If
there is chaos for real fractional order at a certain parameter
value, the trajectories blow up and the system is no longer bounded if
we turn on complex order. We will demonstrate these findings on a case-by-case basis below.
The extension to complex orders for discontinuous systems is not straightforward since the 
underlying variables become complex and effectively two-dimensional.
We have chosen certain rules for extending the map to complex order. However,
other generalizations are possible. 

a) Gauss map: The Gauss map is defined as
$$x_{n+1}=\exp{(-\nu x^{2}_{n})}+\beta,$$ 
where $\nu$ and $\beta$ are the parameters. The parameter  $\nu$ is
fixed and $\nu=7.5$. 
The parameter $\beta$ lies between [-1,1]\\ 
b) Logistic map: 
This  map is given as $$x_{n+1}=\lambda x_{n}(1-x_{n}),$$
where $\lambda$ is a parameter that lies between [0,4]. \\
c)Circle map: The
circle map is given as 
$$x_{n+1}=x_{n}+\Omega-\frac{K}{2\pi} \sin(2 \pi x_{n})\vert(mod\;1).$$
Here $x_{n+1}$ is computed ($mod\;1$) and $K$ is a constant. 
We fix the value of $\Omega=0.6$ for the rest of the paper. 
In our simulations, $K$ lies between [0,5.5].\\
d)Bernoulli map: 
The Bernoulli map is defined as $$x_{n+1}=px _{n}\vert(mod\;1).$$
Here, $p$ lies between [0,2]\\ 
e)Tent map is a piecewise linear map defined  as
\begin{equation}
	x_{n+1}=
	\begin{cases}
		\mu x_{n},         & \vert x_n \vert < \frac{1}{2}, \nonumber \\ 
		\mu (1-x_{n}),     & \vert x_n \vert > \frac{1}{2}. \nonumber
	\end{cases}
\end{equation}
In this map,
the absolute value of the local slope is 
always $\mu$ and it lies between [0,2].

In integer-order difference equations, the 
dynamics of two-dimensional maps is much richer than one-dimensional
maps.  Lorenz system was one of the earliest systems of differential equations
where  chaos was seen \cite{lorenz1963deterministic}.  H{\'e}non introduced a simple 
two-dimensional map that showed similar characteristics. 
The H{\'e}non map is a two-dimensional invertible iterated map 
with squared nonlinearity and strange attractor chaotic solutions. 
In 1976, Michel H{\'e}non, \cite{henon1976two}, introduced this map given as
$$x_{n+1}=1+y_{n}-ax^{2}_{n}$$
$$y_{n+1}=bx_{n}.$$ 
This is one of the earliest and most studied maps \cite{elhadj2013lozi}. 
It has a contraction rate that is independent of the values of variables. 
It reduces to a well-known logistic map for $b=0$. 
Over a  certain range of parameter values, it has bounded solutions 
and it shows chaos at some of the values.
H{\'e}non carried out
numerical experiments and obtained a strange attractor for $a=1.4$ and $b=0.3$. 
However, most of the properties are known only numerically, and hence
Lozi introduced a new map that is hyperbolic, ergodic, and easier to study 
analytically in 1978 \cite{lozi1978attracteur}. 
If the  quadratic term in the H{\'e}non map is replaced by 
the term $|x|$, we get the Lozi map.  Thus, the Lozi map is given as
$$x_{n+1}=1+y_{n}-a|x_{n}|$$
$$y_{n+1}=bx_{n}.$$

The above maps are prototypical and well-studied in two dimensions. 
Naturally, extensions of these maps to
fractional real order have been studied. There are more than
one ways to introduce fractional maps. The fractional H{\'e}non 
map was introduced in 2010 by Tarasov (See Ch. 1 of \cite{luo2011long})
and was later studied in \cite{liu2014discrete,hu2014discrete}.
 Fractional H{\'e}non map in 
$3-d$ \cite{jouini2019fractional}, and chaotic synchronization in fractional 
H{\'e}non map \cite{liu2016chaotic}
has also been investigated. Fractional Lozi map is studied in 
\cite{khennaoui2019fractional} and 
synchronization of fractional-order Lozi maps is 
considered in \cite{megherbi2017new}.
We extend these studies to complex fractional-order. 

\section{Definition of Fractional maps}
We study the maps mentioned above 
in complex fractional-order by using the definition 
of the fractional difference operator 
introduced in \cite{miller1988fractional,atici2010modeling} later extended to 
complex order \cite{bhalekar2022stability}. The definition for the 
$1d$ 
nonlinear map is given as,
\begin{equation}
	x(t)=x(0)+\frac{1}{\Gamma(\alpha)}\sum_{j=1}^{t}\frac{\Gamma(t-j+\alpha)}{\Gamma(t-j+1)}[f(x(j-1))-x(j-1)], \label{eq1}
\end{equation}
where, $f(x)$ is a nonlinear map as defined in a) to e). (The reason for 
subtracting the term $x(j-1)$ in the RHS of the above equation is
to connect the discrete dynamical systems with difference
equations as discussed in \cite{deshpande2016chaos}).

While extending the circle and Bernoulli map to the complex domain, we
define the modulo function as follows. If $x=r\exp(\iota\theta)$ and
$r>1$, we set $r_{new}=r-int[r]$  and $x=r_{new}\exp(\iota\theta)$. 

For $2d$ maps, there
are two possibilities.
The H{\'e}non map which is given by $x_{n+1}=1-ax_n^2+y_n$; $y_{n+1}=bx_n$ or
an equivalent formulation is $x_{n+1}=1-ax_n^2+bx_{n-1}$. These two
formulations are equivalent for integer-order maps. However, extending
them to fractional-order leads to expressions that are not equivalent to
each other. In the first case, we can 
formulate the fractional-order system as
\begin{eqnarray}
	x(t)&=&x(0)+\frac{1}{\Gamma(\alpha)}\sum_{j=1}^{t} \label{henon2d}
	\frac{\Gamma(t-j+\alpha)}{\Gamma(t-j+1)}\nonumber\\
	&&\times[1+y(j-1)-ax^{2}(j-1)-x(j-1)], \nonumber \\
	y(t)&=&y(0)+\frac{1}{\Gamma(\alpha)}\sum_{j=1}^{t}\frac{\Gamma(t-j+\alpha)}{\Gamma(t-j+1)}\nonumber\\ 
	&&\times[bx(j-1)-y(j-1)]. 
\end{eqnarray}
We denote this model as H2.
In the second case, we can formulate
\begin{eqnarray}
	x(t)=x(0)+\frac{1}{\Gamma(\alpha)}\sum_{j=1}^{t}\frac{\Gamma(t-j+\alpha)}{\Gamma(t-j+1)} \nonumber \\	
	\times \left[1+bx(j-2)-ax^{2}(j-1)-x(j-1)\right]. \label{henon1d}
\end{eqnarray}
The formulation involving
delay is used in \cite{liu2014discrete}.
We denote this model as H1.
Some authors use the first definition
\cite{hu2014discrete}. 
(In fact, for the first definition, we can use different 
orders $\alpha$ and $\beta$ for the variables $x$ and $y$. 
However, we do not deal with such generalization in this
work.)
We could choose either formulation. The 
second formulation is computationally
more efficient since it involves only one variable and could be 
a prototype for the 
delay system. We have carried out studies using both
definitions in this work. We note that our major 
conclusions  are not affected by
this choice.

Similarly, we can formulate a fractional-order Lozi system either as
\begin{eqnarray}
	x(t)&=&x(0)+\frac{1}{\Gamma(\alpha)}\sum_{j=1}^{t}
	\frac{\Gamma(t-j+\alpha)}{\Gamma(t-j+1)} \nonumber\\
	&&	\times[1+y(j-1)-a\vert x(j-1)\vert -x(j-1)] \nonumber\\
	y(t)&=&y(0)+\frac{1}{\Gamma(\alpha)}\sum_{j=1}^{t}
	\frac{\Gamma(t-j+\alpha)}{\Gamma(t-j+1)} \nonumber\\
	&&	\times[bx(j-1)-y(j-1)], 
	\label{lozi2d}	
\end{eqnarray}
or
\begin{eqnarray}
	x(t)=x(0)+\frac{1}{\Gamma(\alpha)}\sum_{j=1}^{t}\frac{\Gamma(t-j+\alpha)}{\Gamma(t-j+1)} \nonumber \\
	\times \left[1+bx(j-2)-a\vert x(j-1)\vert -x(j-1)\right], \label{lozi1d}
\end{eqnarray}
where $\alpha$ is the order of the fractional difference operator. In our case,
$\alpha$ is a complex number with $0<Re(\alpha)<1$. 
We denote models defined by eq(\ref{lozi2d}) and (\ref{lozi1d}) by model L2 and 
L1 respectively.

To systematically investigate the impact of complex order on
fractional maps, we set $\alpha=\alpha_0e^{i r\pi/2}$ with
$0<\alpha_0<1$ and $0\le r<1$. It reduces to real fractional
order for $r=0$.

\section{Dynamics and bifurcations for fractional-order maps}

\begin{figure*}[ht!]
	\subfloat[]{%
		\centering\includegraphics[width=3.4in,height=2.7in]{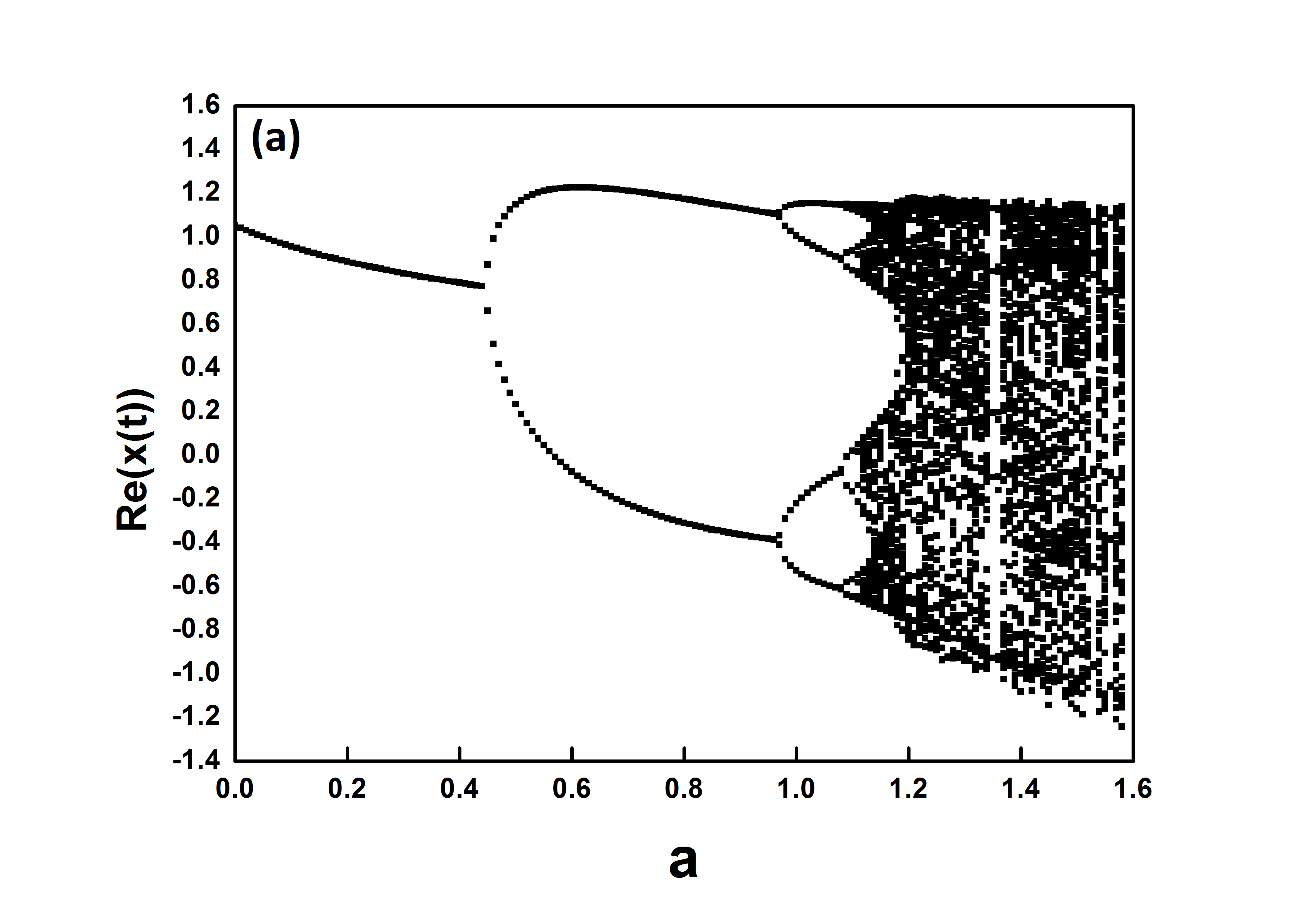}
	}
	\subfloat[]{%
		\centering\includegraphics[width=3.4in,height=2.7in]{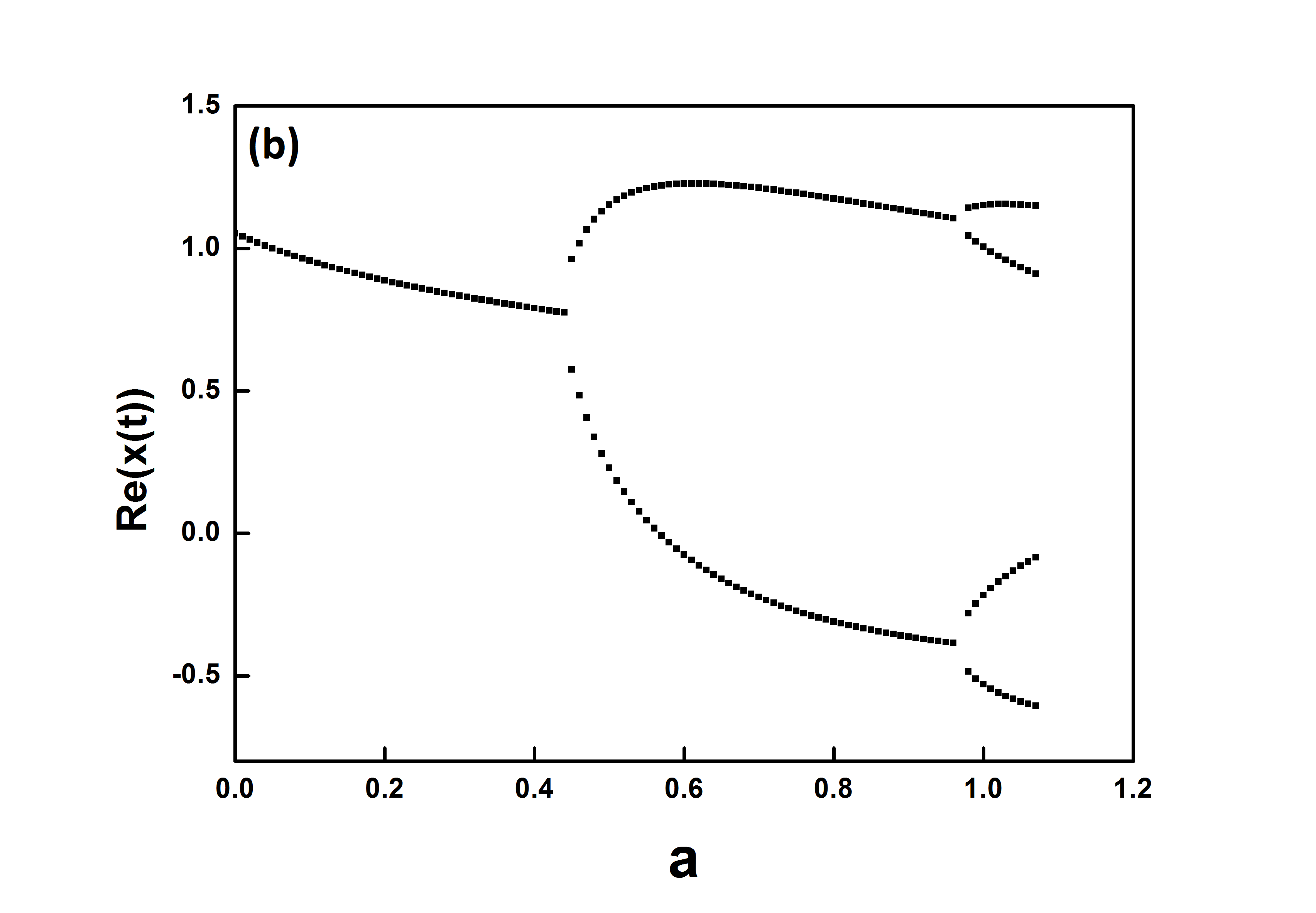}
	}\\
	\subfloat[]{%
		\centering\includegraphics[width=3.4in,height=2.7in]{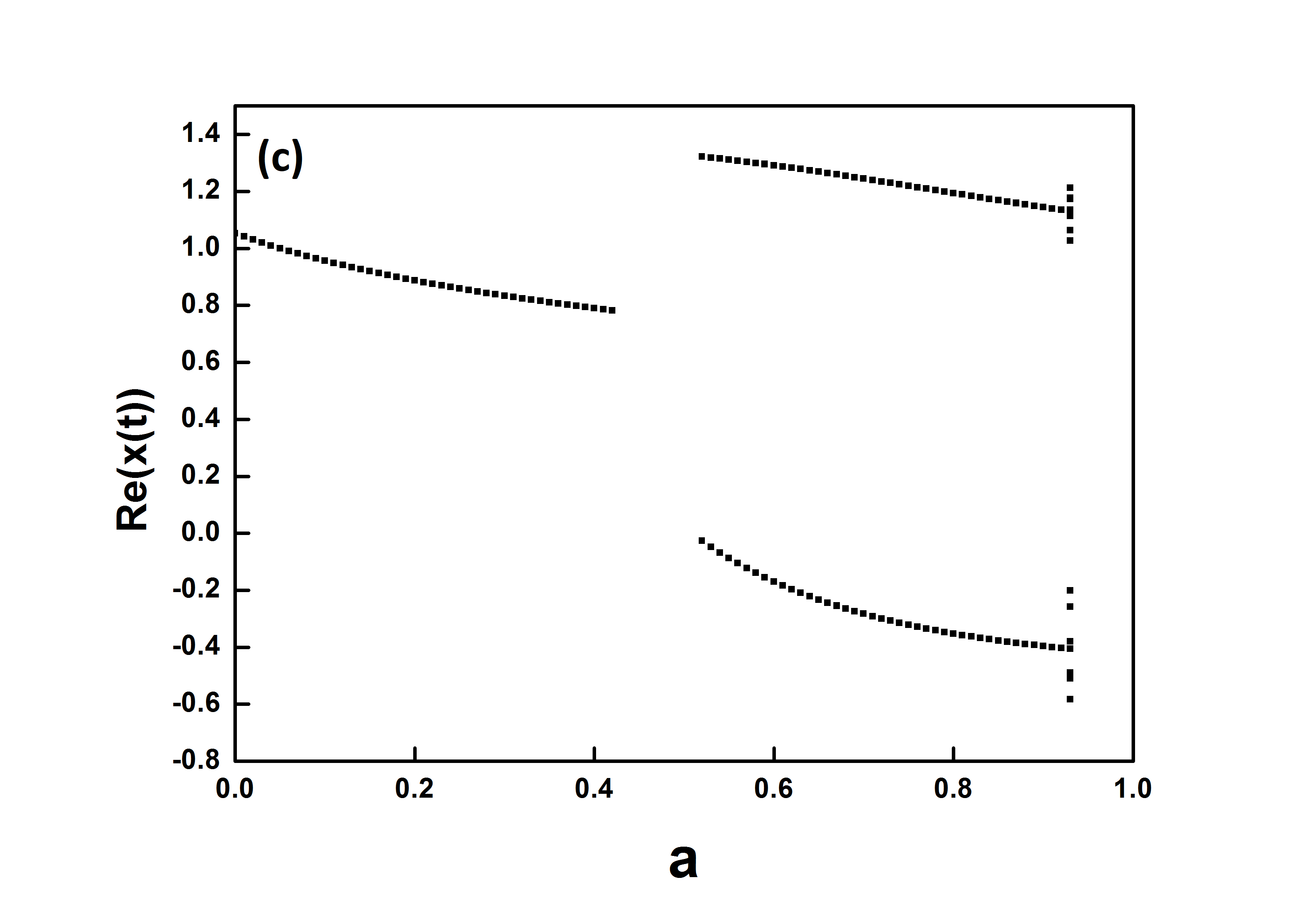}
	}
	\subfloat[]{%
		\centering\includegraphics[width=3.4in,height=2.7in]{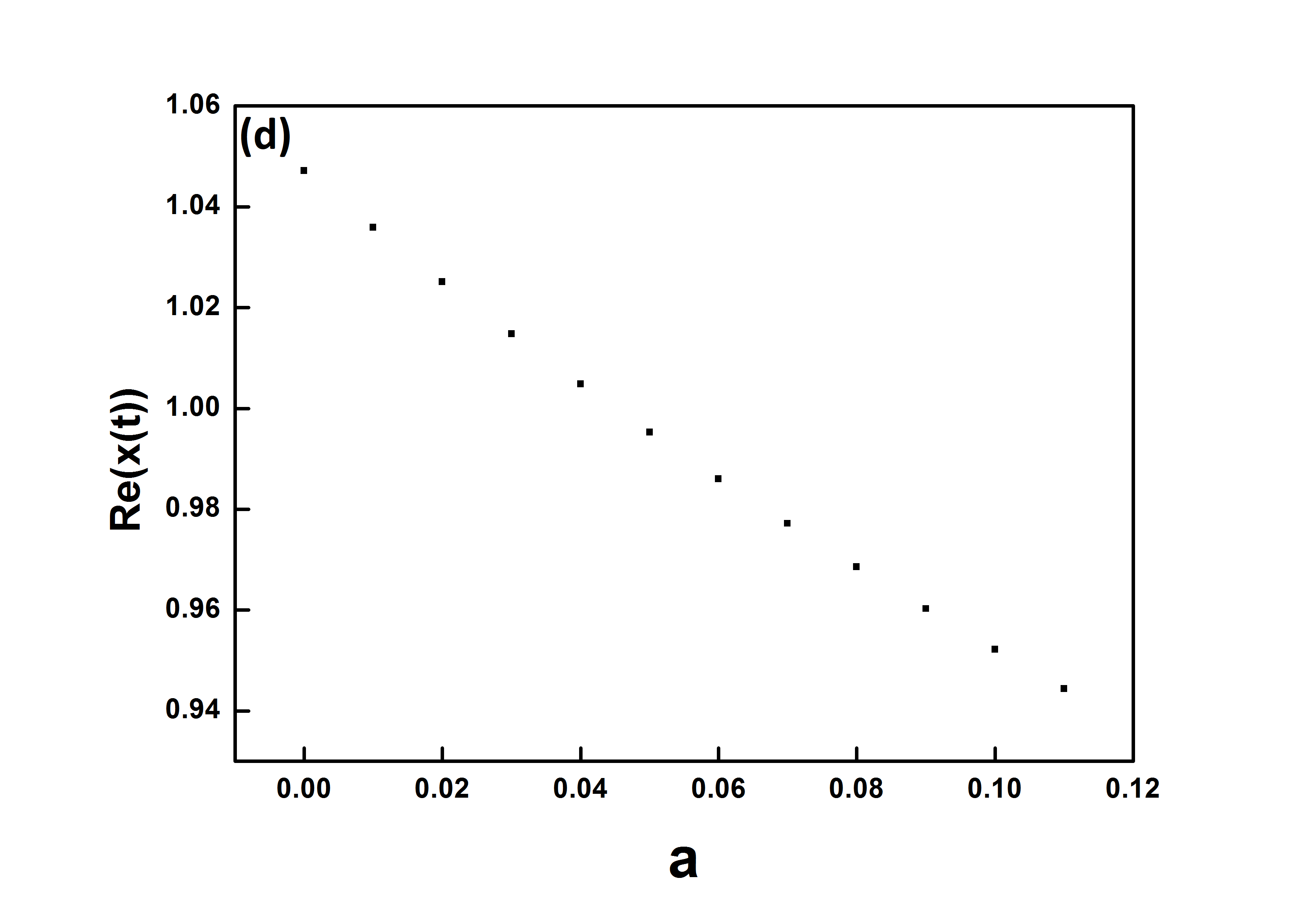}
	}\\
	\centering
	\subfloat[]{%
		\centering\includegraphics[width=3.4in,height=2.7in]{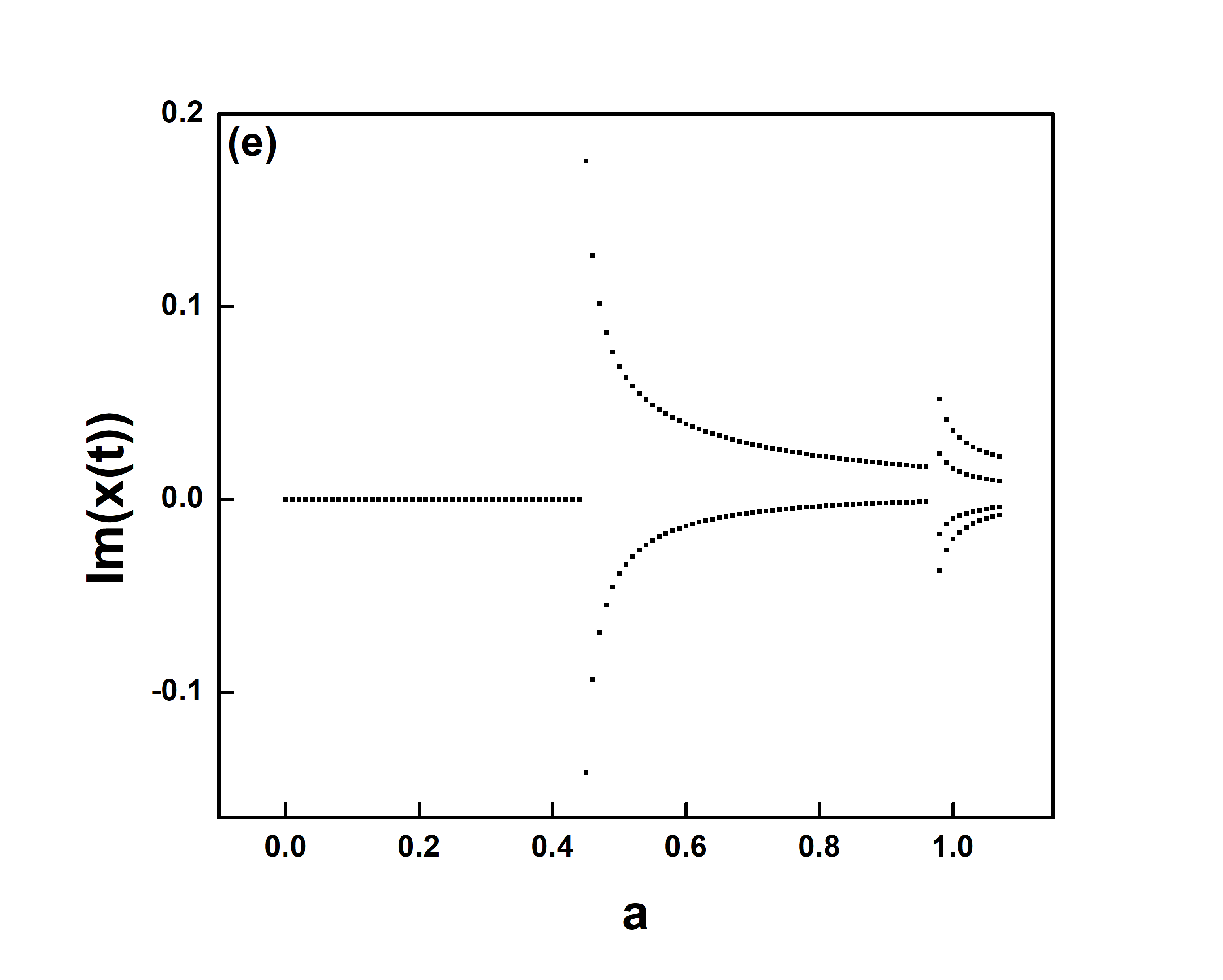}
	}
	\caption{Bifurcation Diagrams($a$ vs $Re(x(t))$) for 
	the H{\'e}non map (model H1)for $b=0.05$ and
	$\alpha_0=0.8$ for (a)$r=0$, (b)$r=0.01$, 
	(c)$r=0.1$, (d)$r=0.5$ and (e)$a$ vs $Im(x(t))$ for 
	$\alpha_0=0.8, r=0.01$.}
	\label{figa}
\end{figure*}

A powerful tool to understand the dynamics is a bifurcation diagram.
The span of variable values 
is clear when we plot  
values of variables after a certain transient. 
At first, we see the bifurcation diagram for the H{\'e}non map
of order $\alpha=\alpha_0e^{i r\pi/2}$ as mentioned above.
We have checked the values $\alpha_0=$ 0.1, 0.2, 0.3, 0.4, 0.5, 0.6, 0.7, 0.8, and 0.9. 
For all these values, we have checked the cases $r=$ 0, 0.01, 0.1, and
0.5 with initial condition $x(0)=y(0)=0$ for model H1. We do not 
observe chaos for $r\ne 0$ indicating that the  chaotic attractor is
destroyed when we introduce complex order. We have checked 
results with a 
few different initial conditions and the
bifurcation
diagram does not change indicating that multistability is not
very pronounced for fractional-order. 

We have shown the bifurcation diagram for $\alpha_0=0.8$ and $r=$ 0, 0.01, 0.1,
and 0.5 in Figure (\ref{figa}). The chaos disappears even
for small values of $r$. We have carried out the same exercise
for model H2 and we do not observe a stable chaotic orbit
with initial conditions $x(0)=y(0)=0$ for the cases mentioned above.

\begin{figure*}[ht!]
	\subfloat[]{%
		\centering\includegraphics[width=3.4in,height=2.7in]{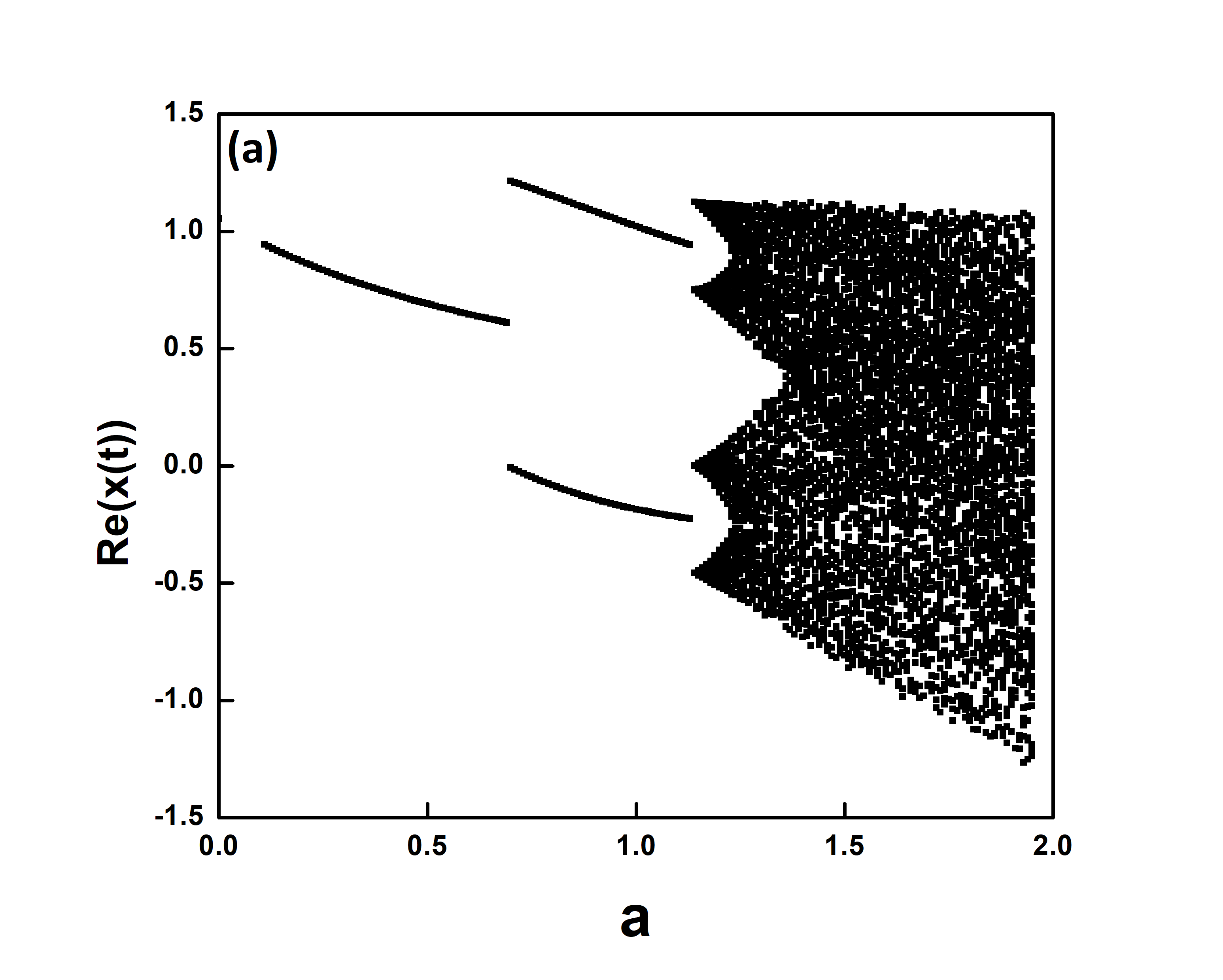}
	}
	\subfloat[]{%
		\centering\includegraphics[width=3.4in,height=2.7in]{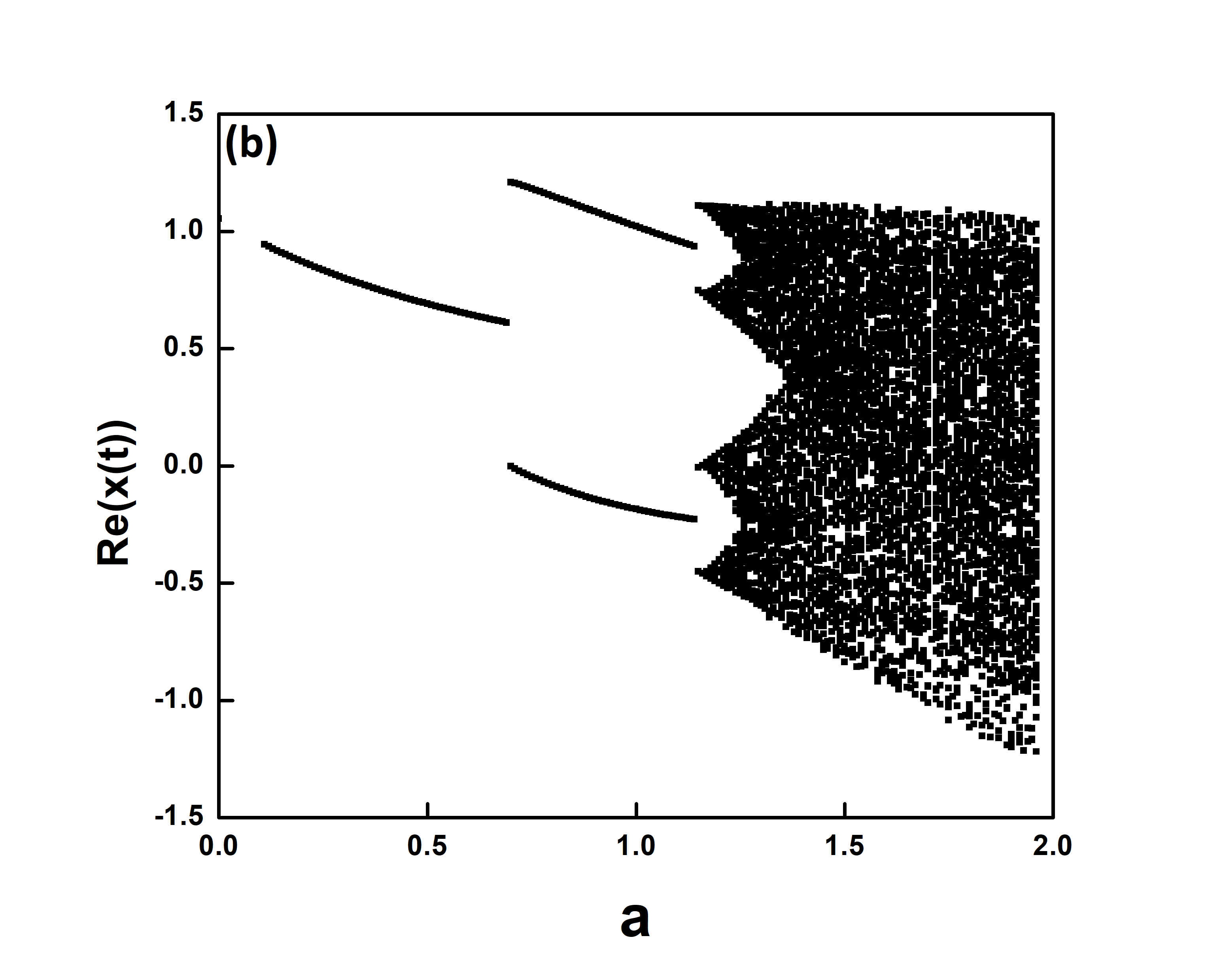}
	}\\
	\subfloat[]{%
		\centering\includegraphics[width=3.4in,height=2.7in]{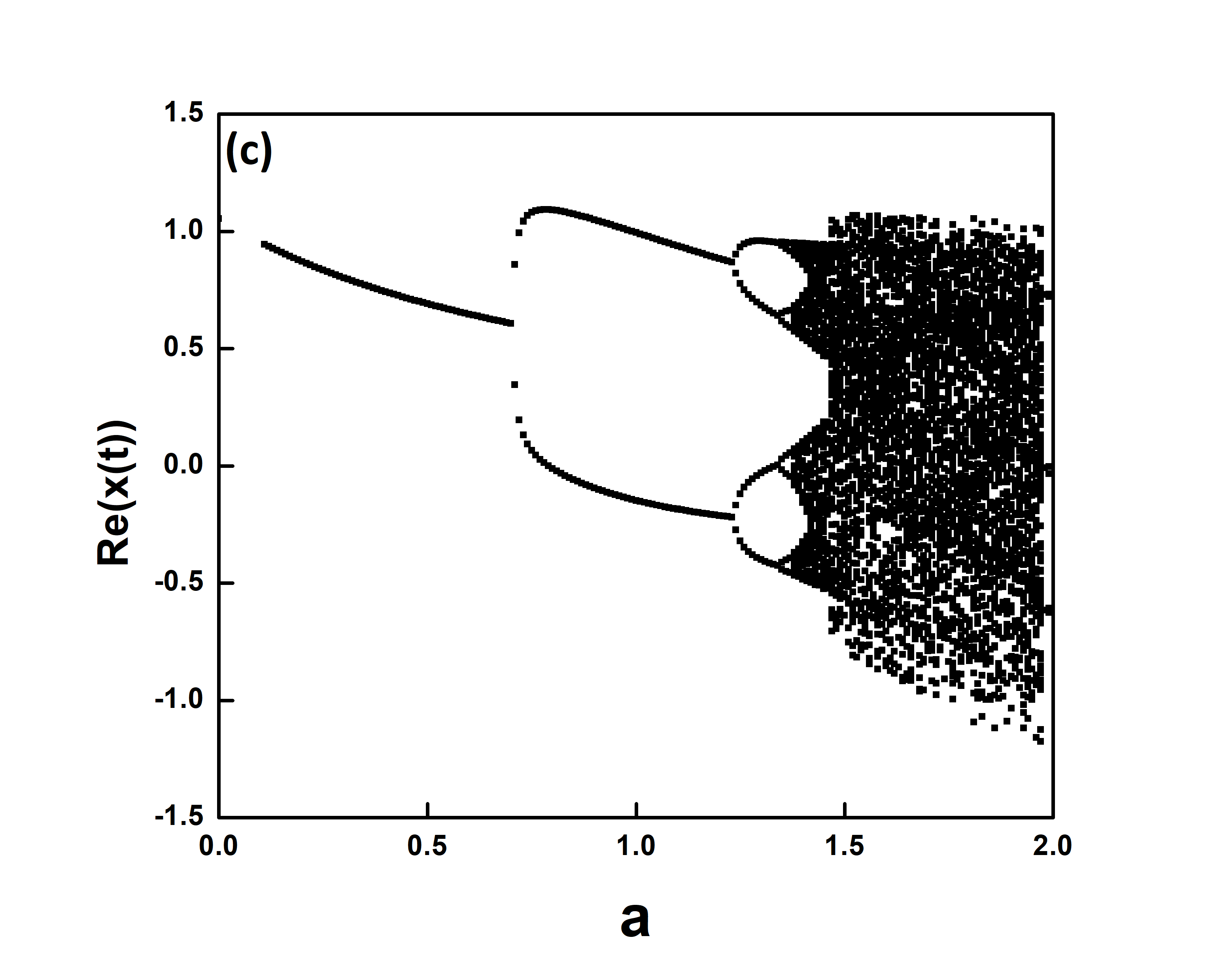}
	}
	\subfloat[]{%
		\centering\includegraphics[width=3.4in,height=2.7in]{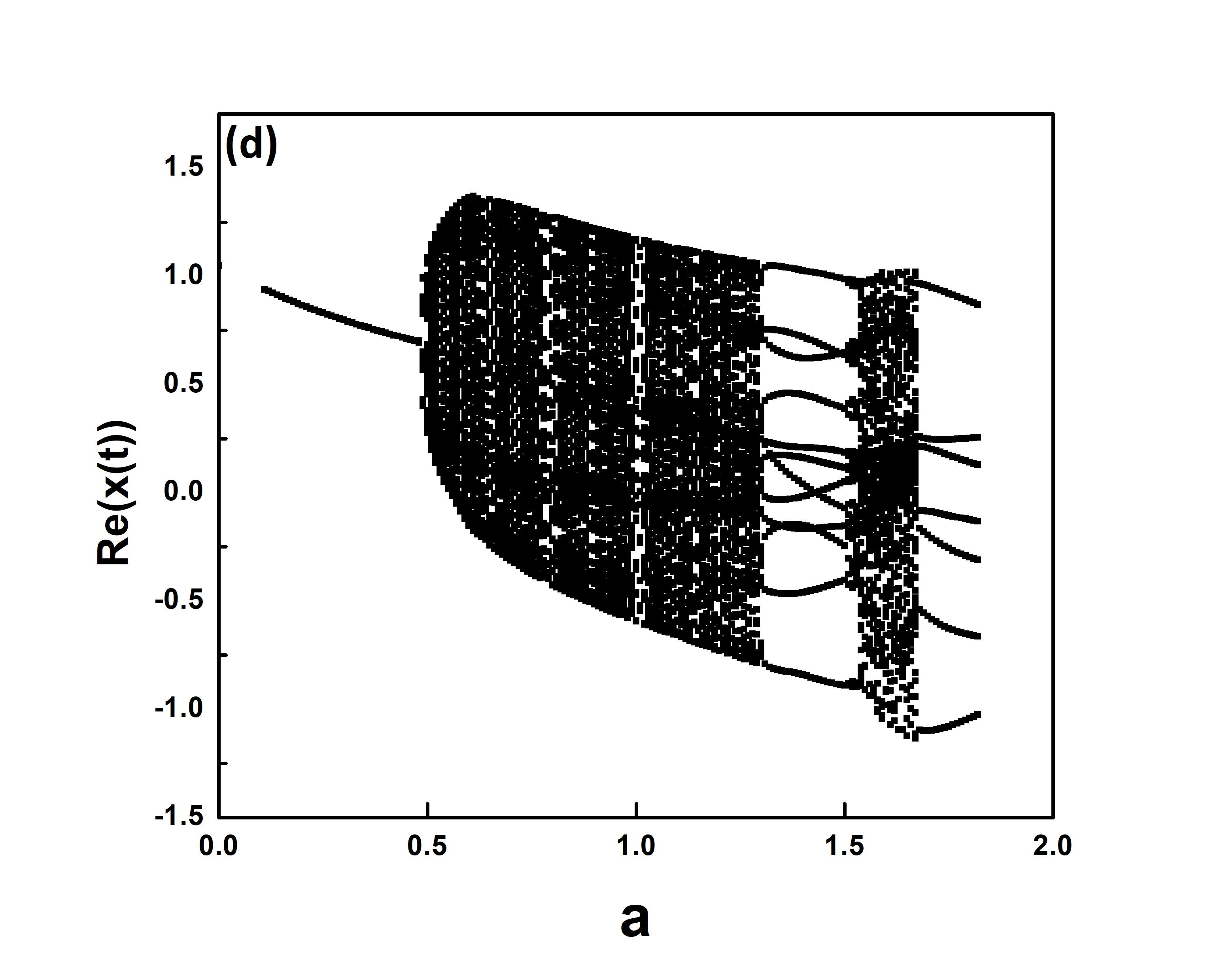}
	}\\
	\centering
	\subfloat[]{%
	\centering\includegraphics[width=3.4in,height=2.7in]{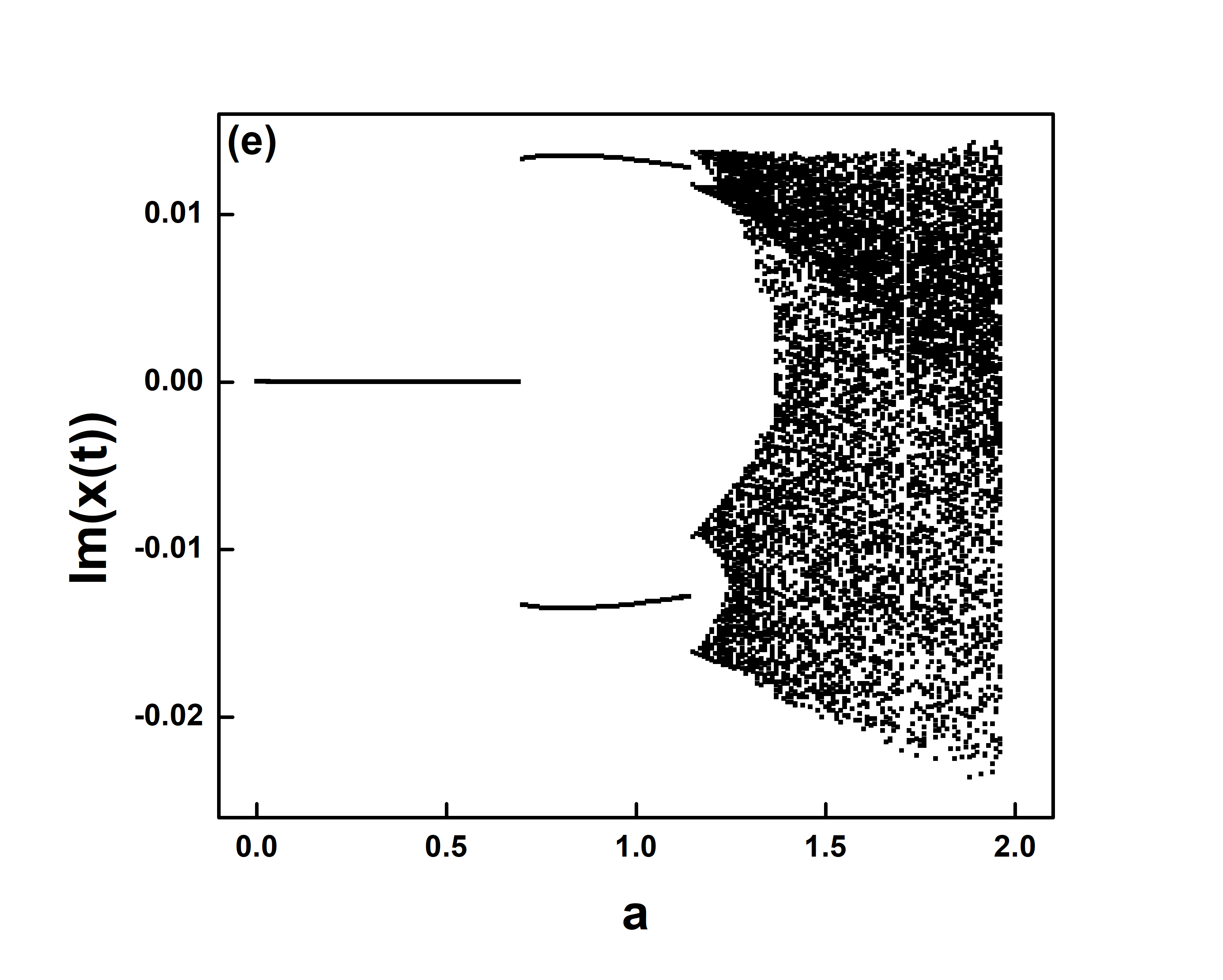}
	}
	\caption{Bifurcation diagrams($a$ vs $Re(x(t))$) for the Lozi map 
	(model L1) for $b=0.05$ and
		$\alpha_0=0.8$ for (a)$r=0$, (b)$r=0.01$, (c)$r=0.1$ and (d)$r=0.5$ and (e)$a$ vs $Im(x(t))$ for $\alpha_0=0.8, r=0.01$.}
	\label{figb}
\end{figure*}

However, the situation is different for
Lozi map. In this case, the chaos
does not disappear.
We have shown the bifurcation 
diagram of the Lozi map for $\alpha_0=0.8$ and $r=$ 0, 0.01, 0.1 and 0.5 for model L1.
The span of variable values indicates that 
there are parameter zones that are chaotic 
or at least periodic with
a very large period (see Figure (\ref{figb})). 
We carry out further 
tests such as finding the Lyapunov exponent to
confirm the presence of chaos.
As mentioned above, the divergence of trajectories is 
exponential in our formulation and computing
Lyapunov exponent is justified.

\begin{figure*}[ht!]
	\subfloat[]{%
		\centering\includegraphics[width=3.4in,height=2.7in]{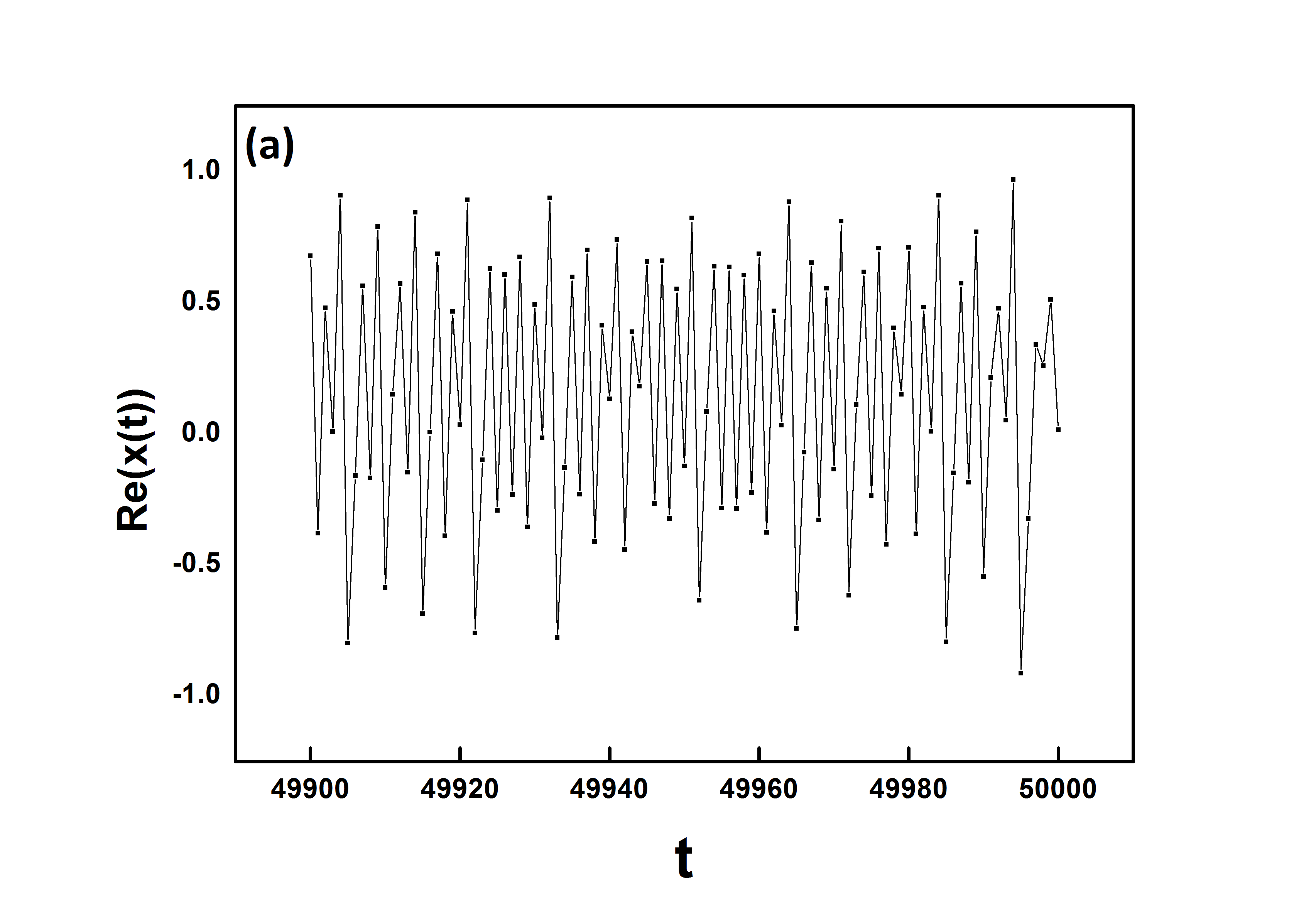}
	}
	\subfloat[]{%
		\centering\includegraphics[width=3.4in,height=2.7in]{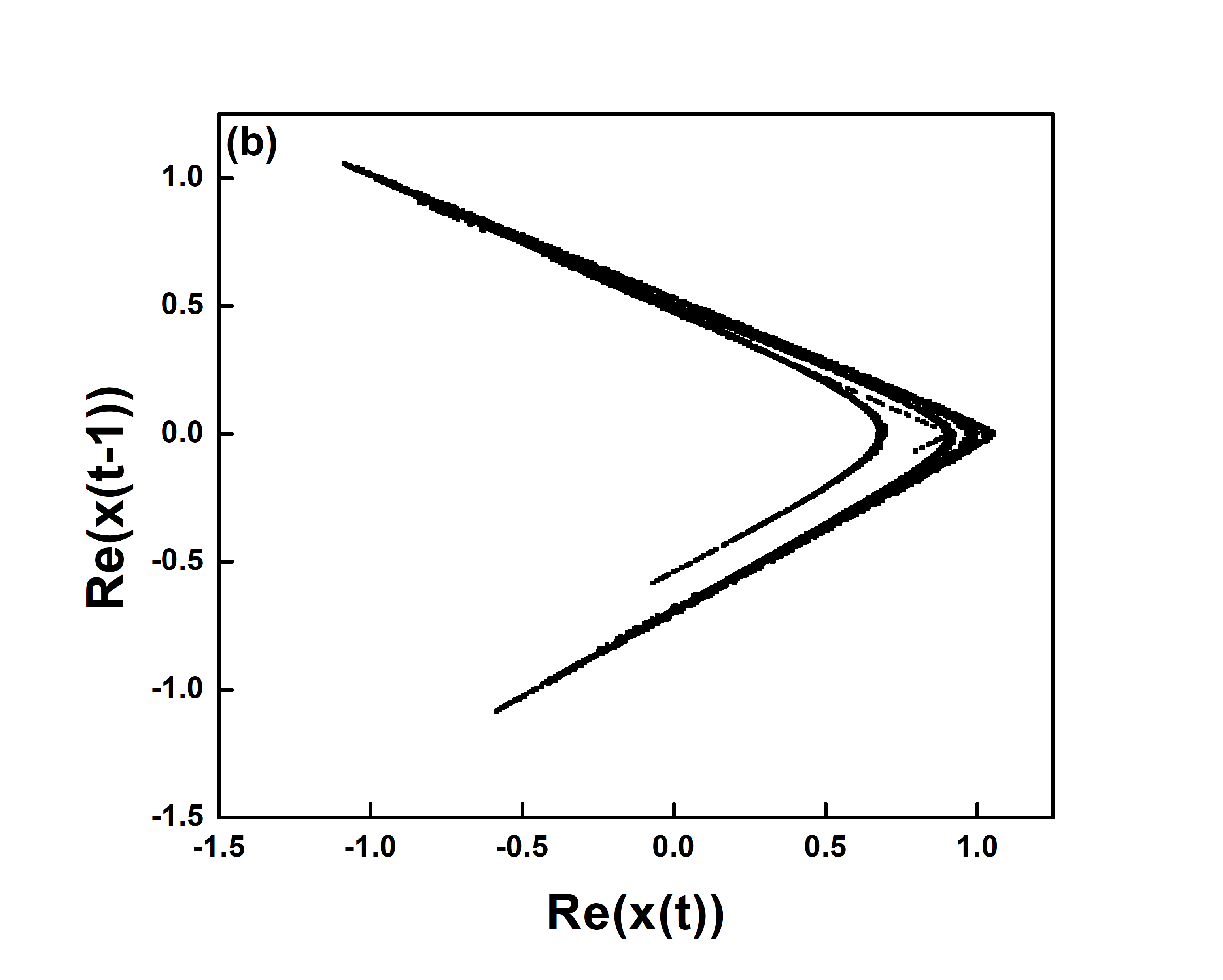}
	}\\
	\subfloat[]{%
		\centering\includegraphics[width=3.4in,height=2.7in]{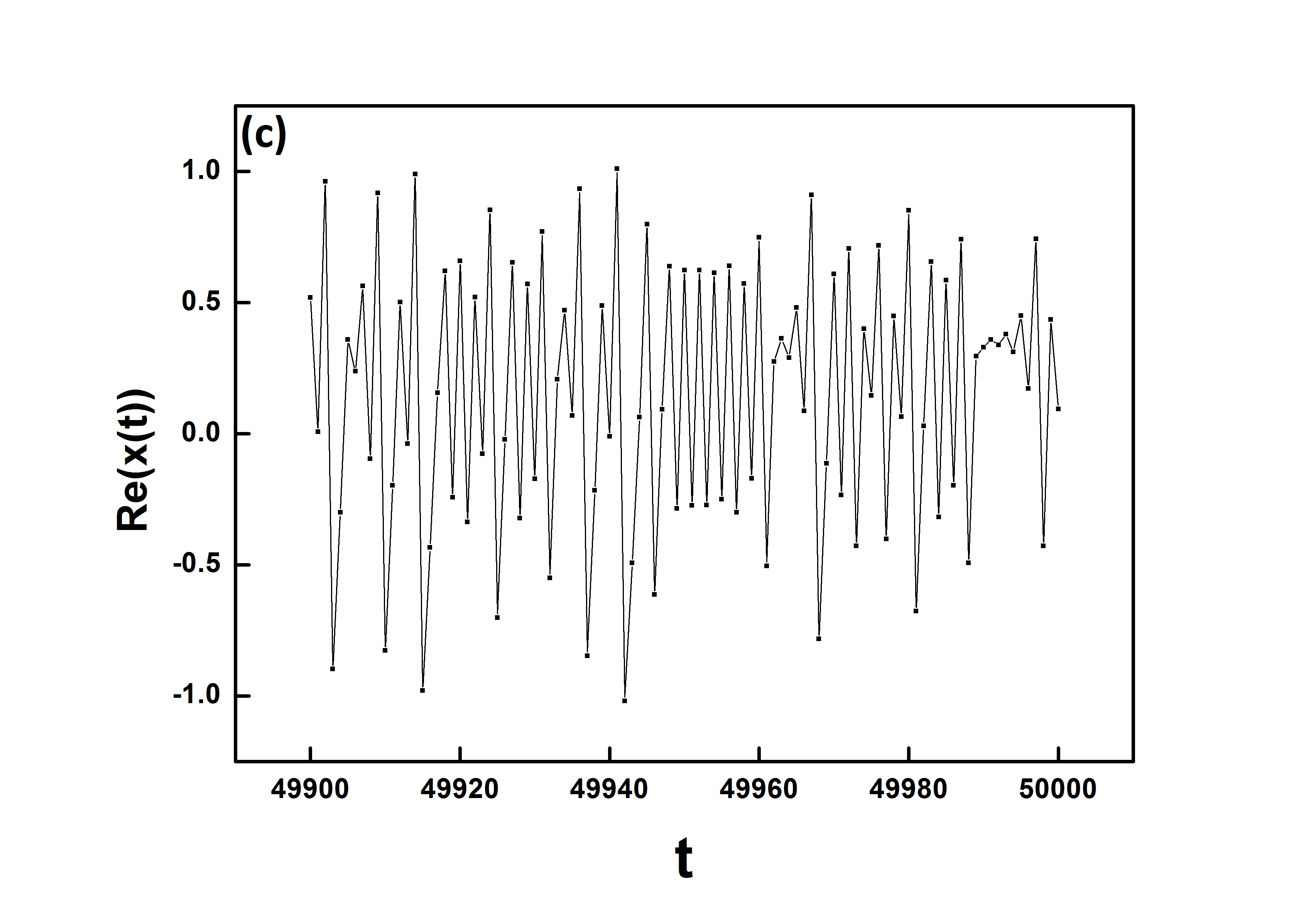}
	}
	\subfloat[]{%
		\centering\includegraphics[width=3.4in,height=2.7in]{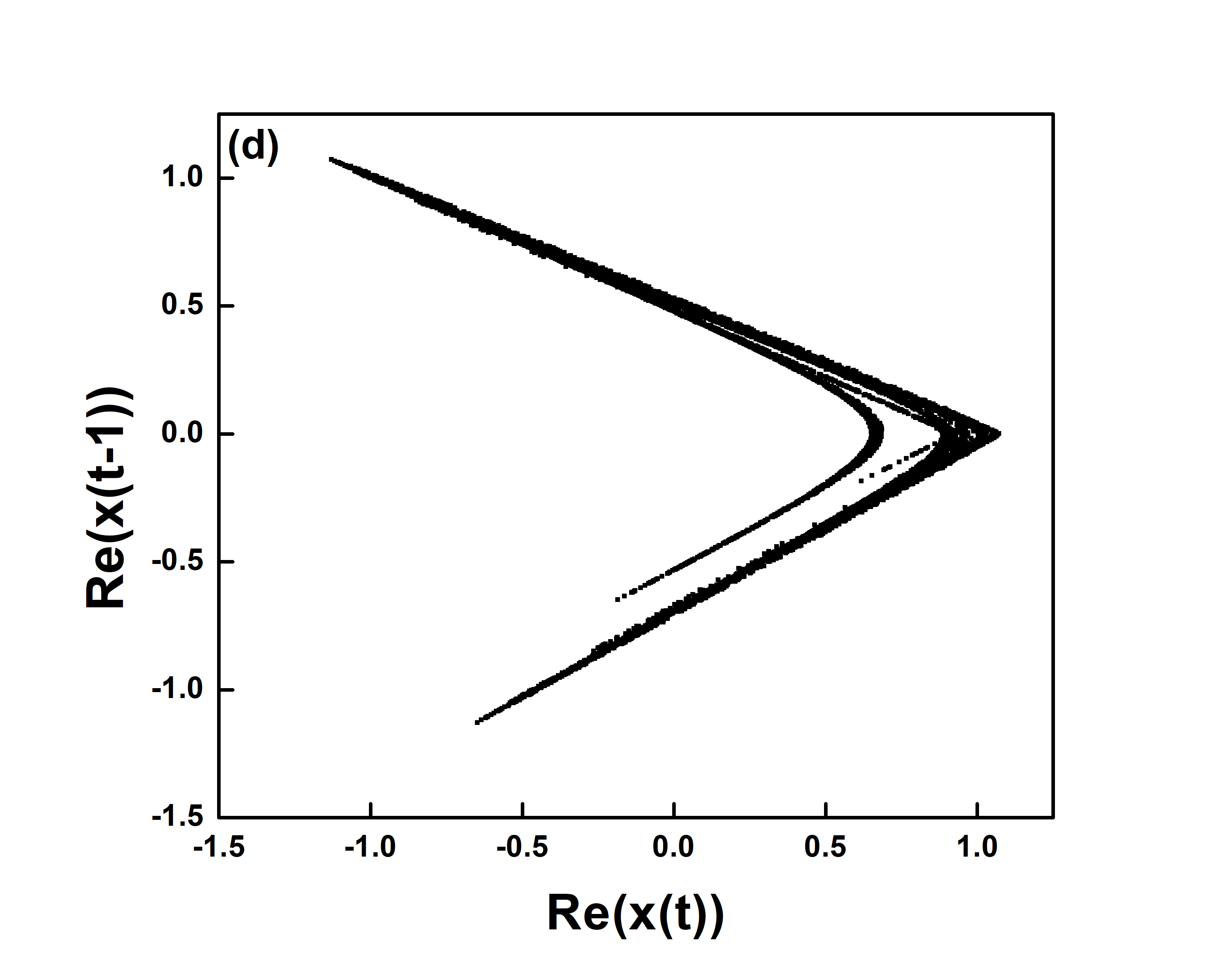}
	}
	\caption{Attractor for Lozi map 
	with $\alpha_0=0.8$, $r=0.1$, $b=0.05$ and 
		$a=1.8$ 
		(a) time series for corresponding to
		model L1 (b) attractor  corresponding to a)
		of model L2 
		(c) time series corresponding to model L2
		(d) attractor corresponding to c).}
	\label{figc}
\end{figure*}

To check that the system is chaotic, we plot time series
for $\alpha_0=0.8$, $r=0.1$, $a=1.8$ for both formulations.
After discarding $5 \times 10^4$ transients, we plot 
the attractor by plotting $x(t)$ versus $x(t-1)$ for the last
$10^4$ time steps. We also plot the time series of the last 100 time steps. 
(see Figure (\ref{figc})).
We observe chaos in the time series, if we carry on the simulations 
for model L2 (eq(\ref{lozi2d})) as well
as model L1 (eq(\ref{lozi1d})) of Lozi maps.
To provide definitive proof that these attractors
are indeed chaotic, we find Lyapunov exponents from their 
time series \cite{wolf1985determining, kodba2004detecting}. 
We use the program for finding the largest Lyapunov exponent from
time series in the above works \cite{lyapmax}.
We find the Lyapunov exponents for both formulations of Lozi maps. 
For the simulation of the Lozi map, 
we obtain the exponent to be 0.264 for model L1
and exponent 
0.326 for model L2 for the parameters mentioned in the caption.
Since both the systems show positive values of the Lyapunov exponent, 
it confirms our prognosis that the Lozi
map shows chaos for the difference equation of complex fractional order. 

\begin{figure*}[ht!]
	\subfloat[Gauss map with $\alpha_0=0.7$ and $r=0.01$.\label{fig40}]{%
		\centering\includegraphics[scale=0.3]{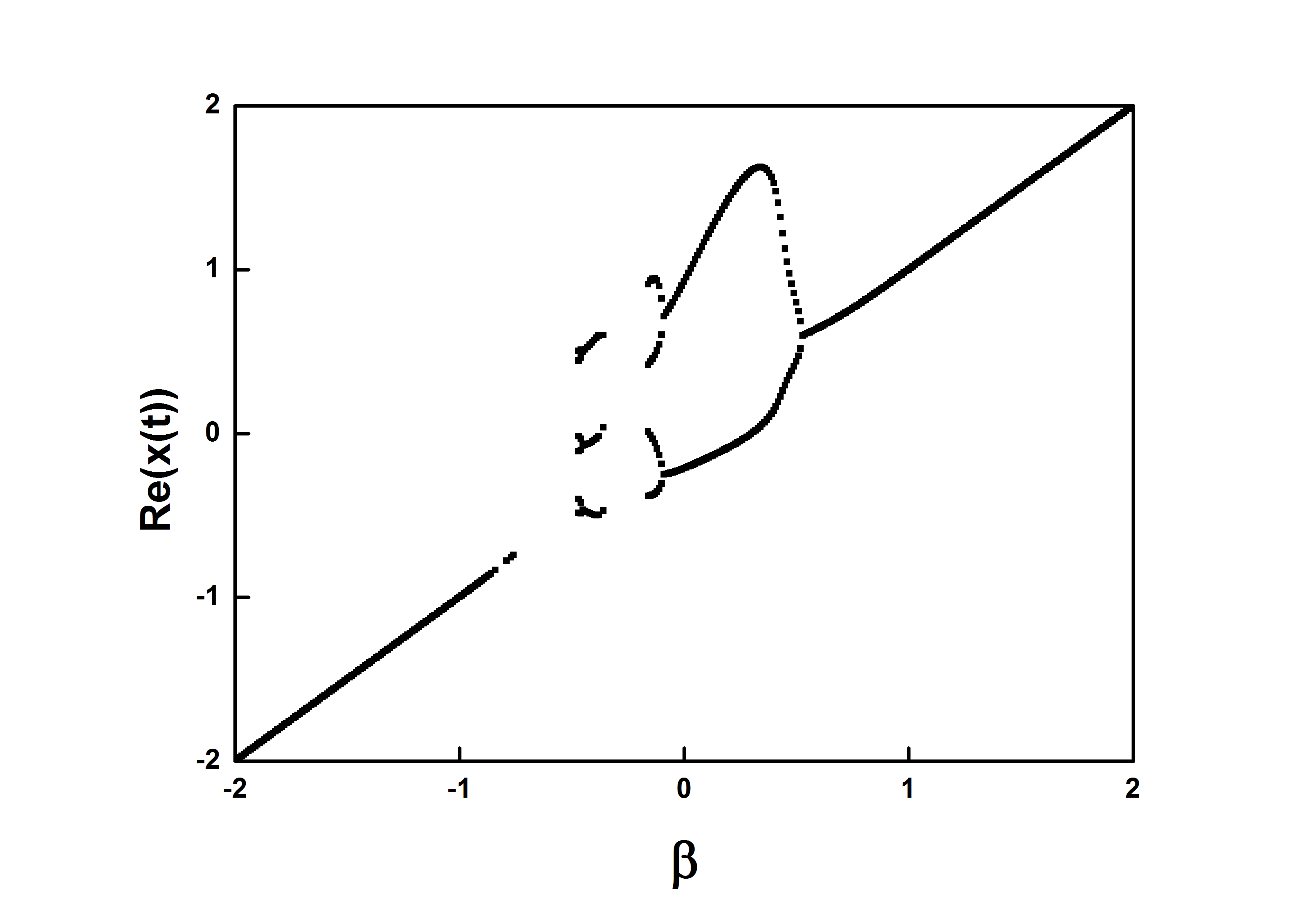}
	}
	\subfloat[Logistic map with $\alpha_0=0.8$ and $r=0.01$.\label{fig41}]{%
		\centering\includegraphics[scale=0.3]{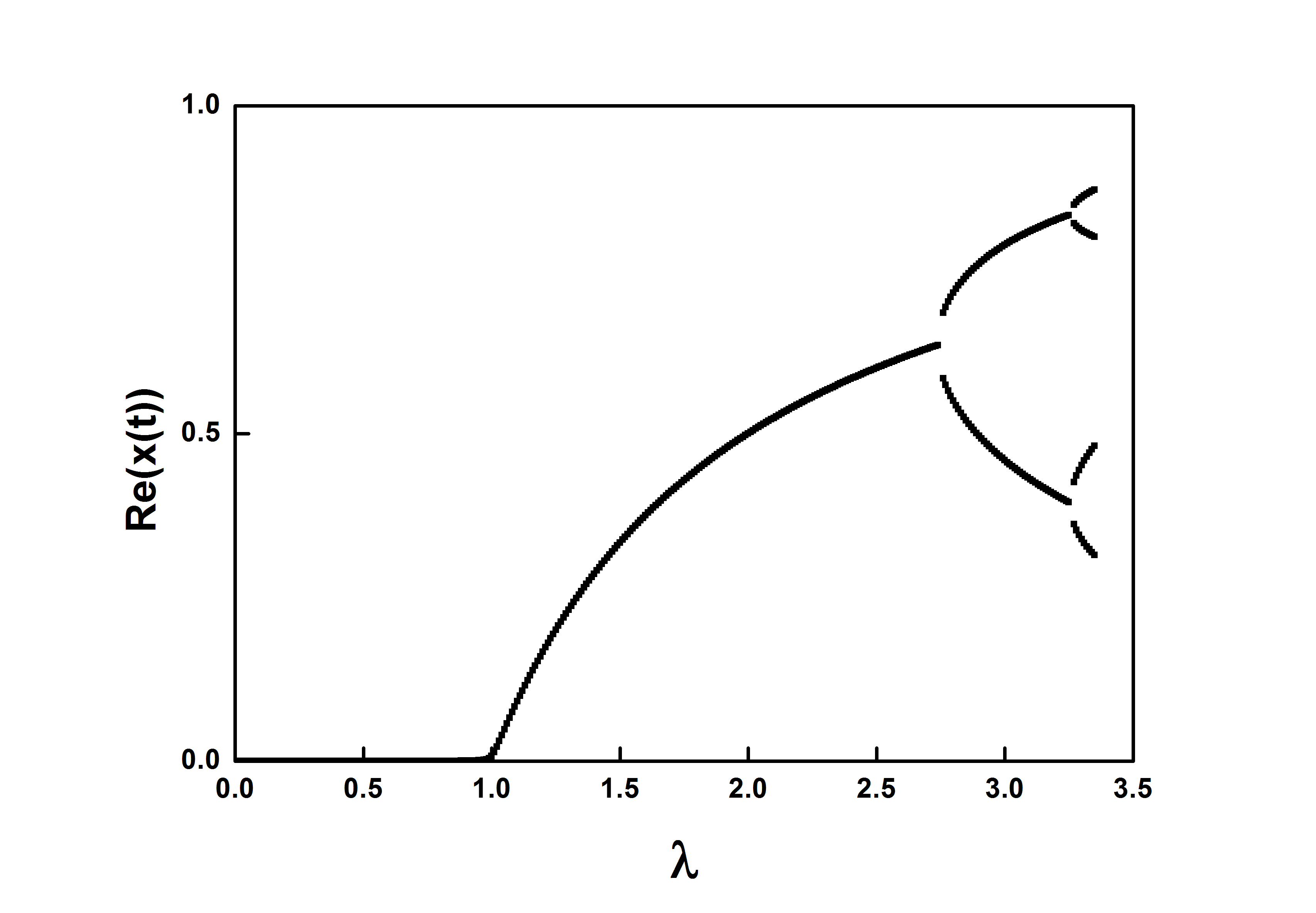}
	}
	\caption{Bifurcation diagrams for Gauss map and logistic map. The bifurcation diagram was observed for the system by taking $10^4$ timesteps and the last 120 values have been plotted. No chaos was observed for these maps.}
	\label{figd}
\end{figure*}

H{\'e}non map is an analytic function and the bifurcation diagram 
(\ref{figa}) does not show chaos for complex fractional-order. 
(The bifurcation diagram is concerning  model H1. However, 
a similar diagram is obtained for model H2.)
The chaotic attractors vanish with the 
slightest introduction of an imaginary part in the order. 
There could be a link between the absence of chaos 
for complex fractional ordered maps 
and the analytic nature of the function maps. 
To investigate if it is indeed so,  we analyze two more 
smooth functions {\it{i.e.}}, continuous and differentiable maps.
We consider 
two of the most popular maps, Gauss and logistic maps. 
Figure (\ref{figd}) shows the bifurcation diagram for Gauss and logistic maps. 
We observe that the chaos vanishes as the order gets slightly complex
as in the case of the H{\'e}non map.  

\begin{figure*}[ht!]
	\centering
	\includegraphics[scale=0.65]{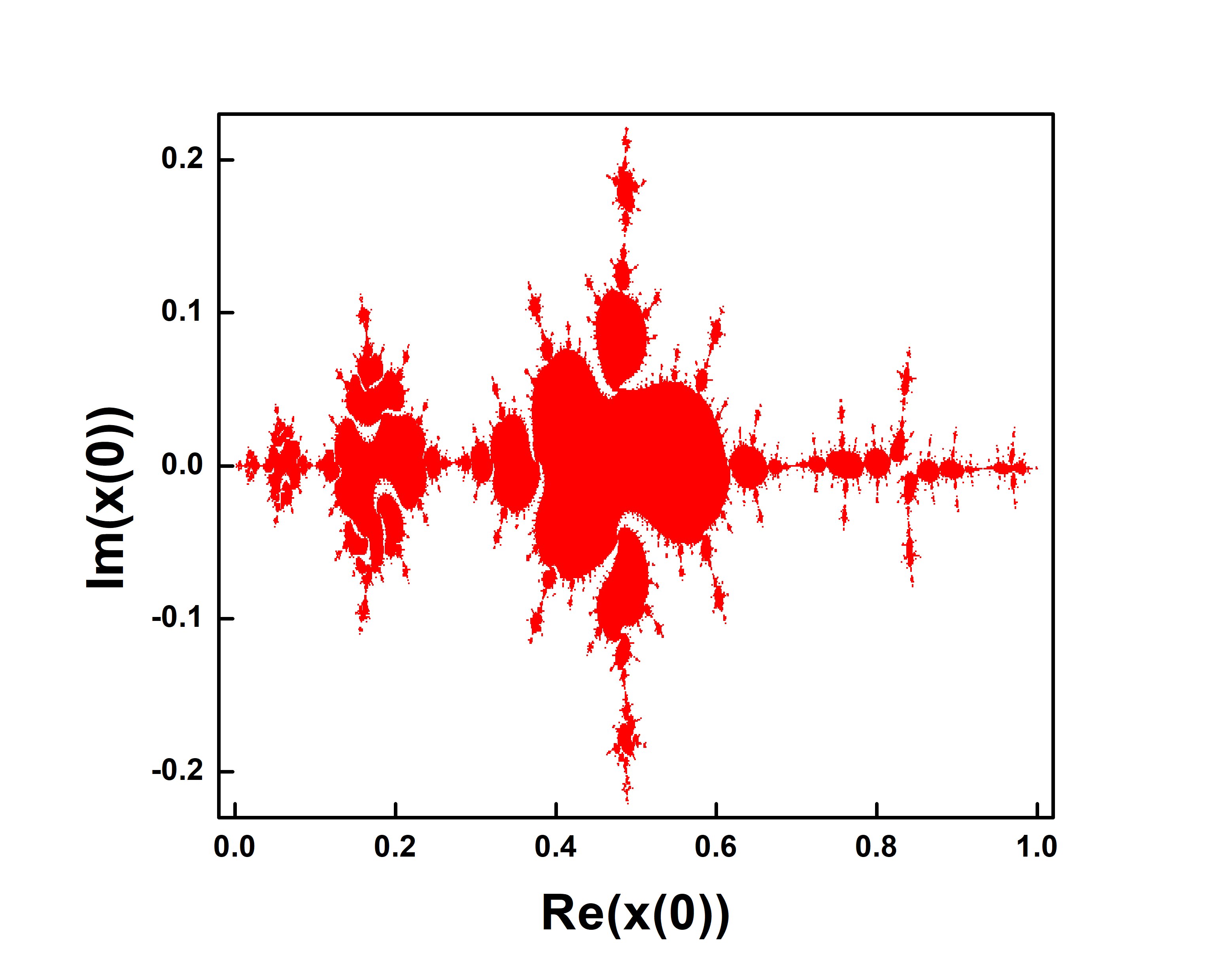}
	\caption{Basin of logistic map with $\alpha_0=0.8$ and $r=0.01$.}
	\label{figd1}
\end{figure*}

\begin{figure*}[ht!]
	\subfloat[H{\'e}non map with $\alpha_0=0.8$, $r=0$, $x_{0}=(0.2,0)$ and $x_{(-1)}=(0.1,0)$.]{
		\includegraphics[width=3.4in,height=2.7in]{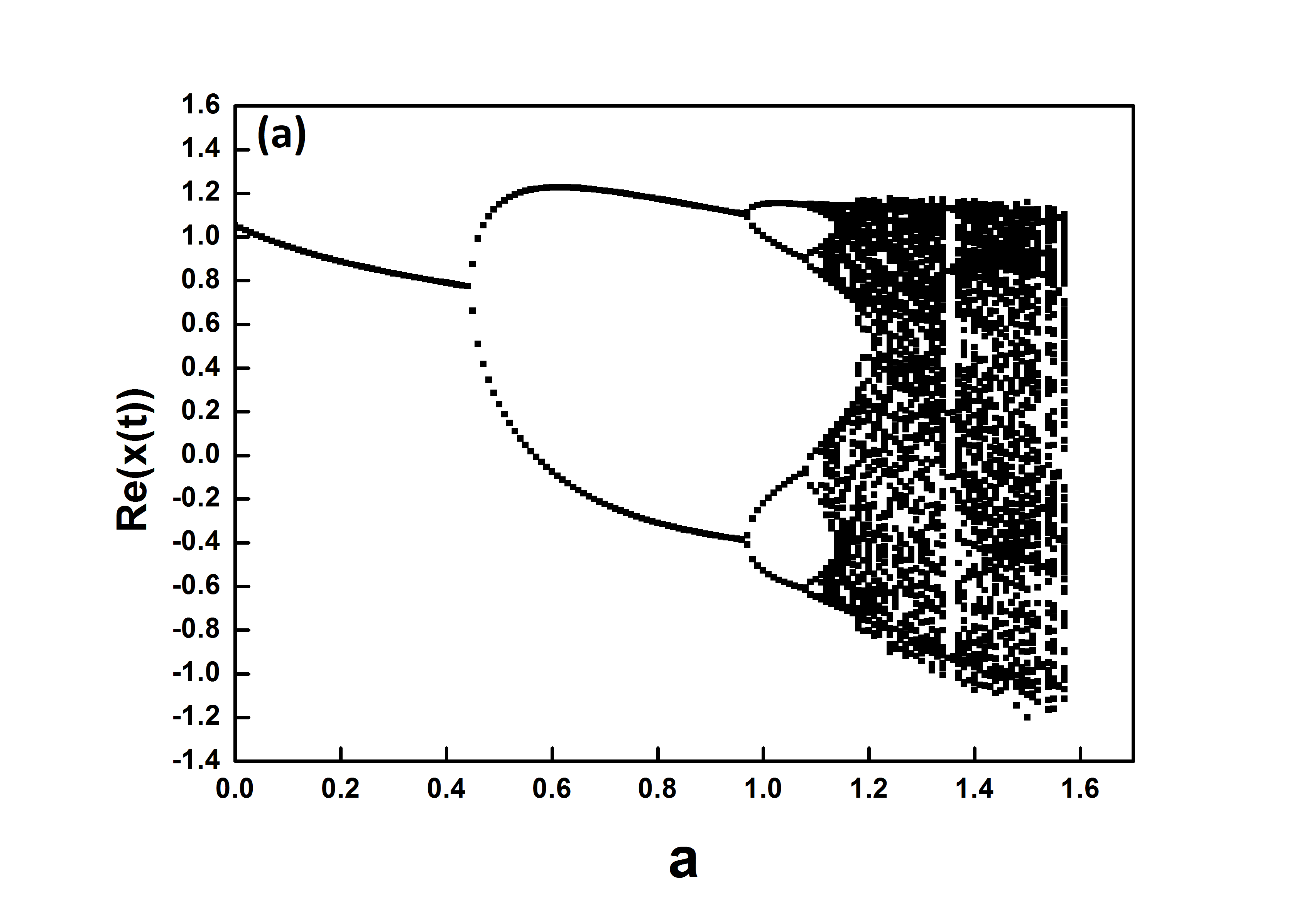}
	}
	\subfloat[ H{\'e}non map with $\alpha_0=0.8$, $r=0$, $x_{0}=(0.2,0.1)$ and $x_{(-1)}=(0.1,0.1)$.]{
		\includegraphics[width=3.4in,height=2.7in]{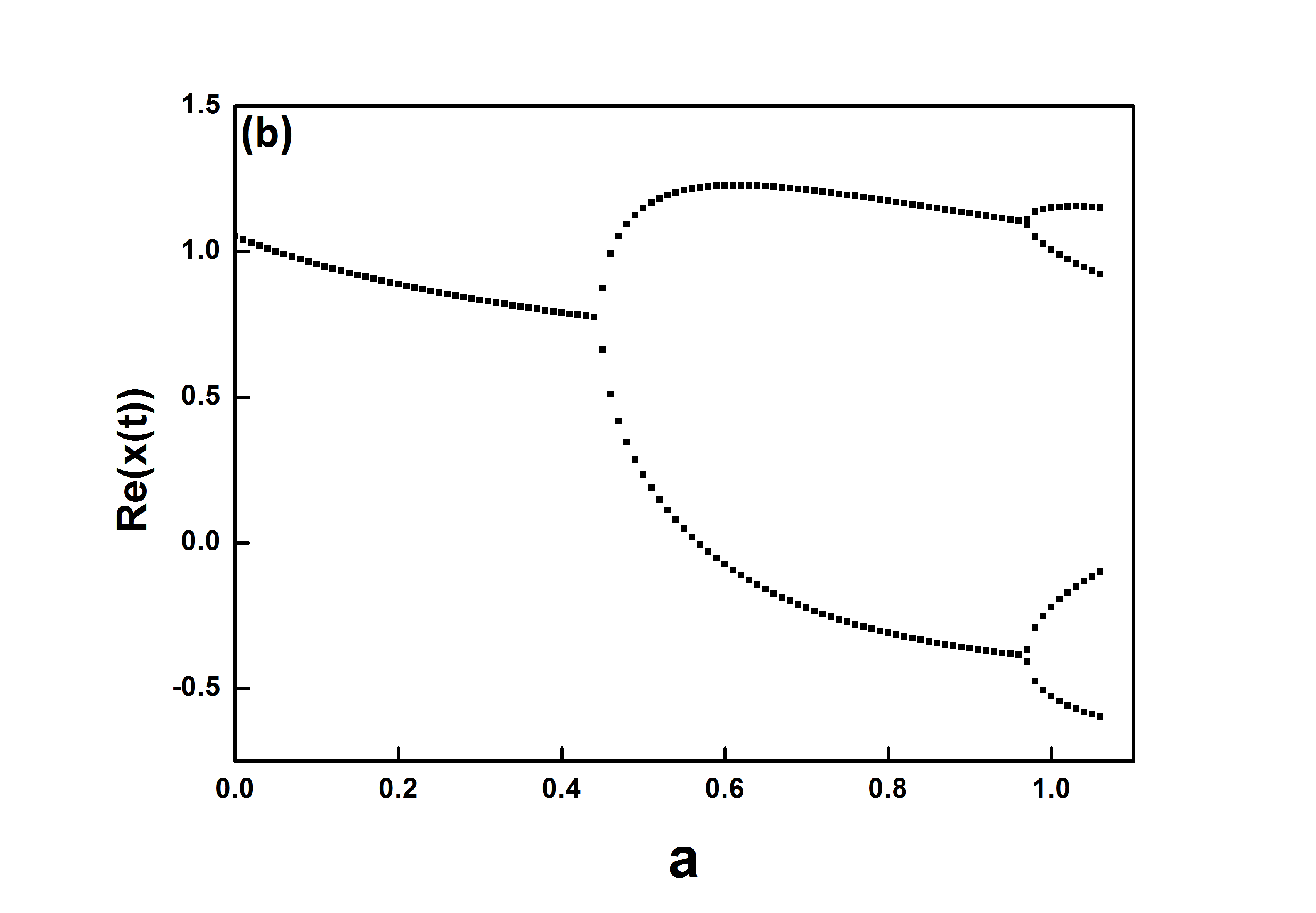}
	}\\
	\subfloat[Gauss map with $\alpha_0=0.9$, $r=0$ and $x_{0}=(0.1,0)$.]{
		\includegraphics[width=3.4in,height=2.7in]{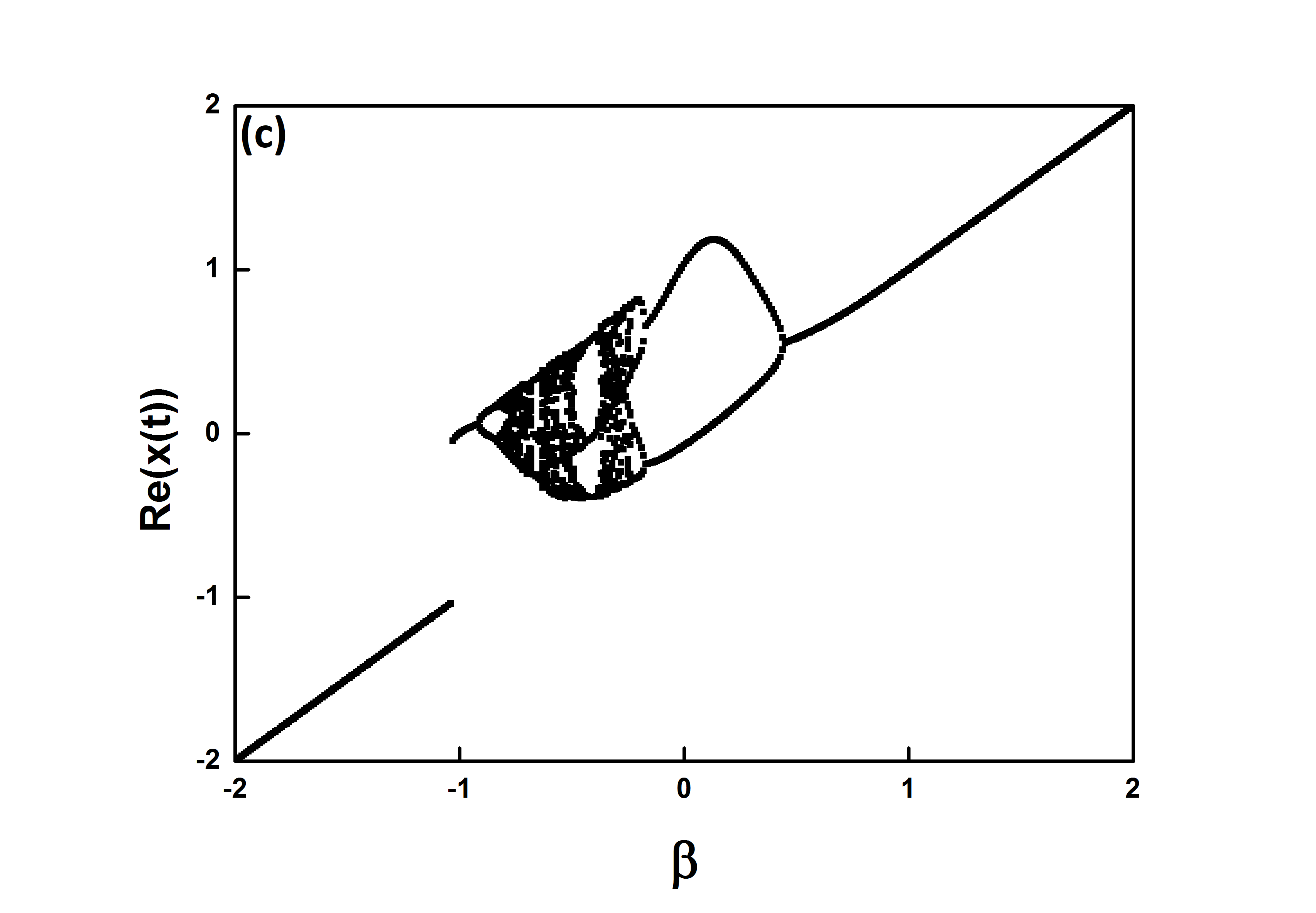}
	}
	\subfloat[Gauss map with $\alpha_0=0.9$, $r=0$ and $x_{0}=(0.1,0.1)$.]{
	\includegraphics[width=3.4in,height=2.7in]{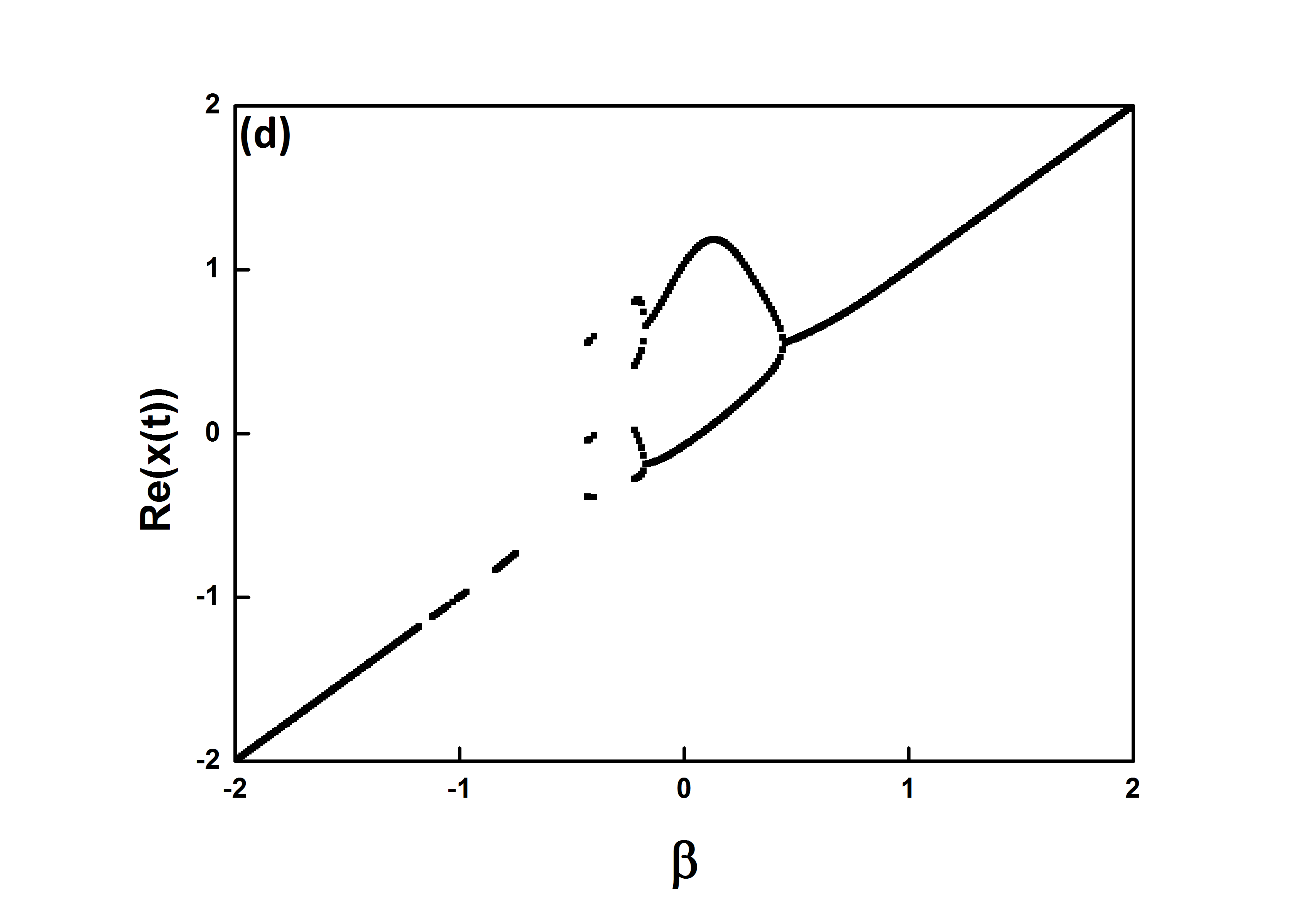}
	}\\
	\subfloat[Logistic map with $\alpha_0=0.2$, $r=0$ and $x_{0}=(0.2,0)$.]{
		\includegraphics[width=3.4in,height=2.7in]{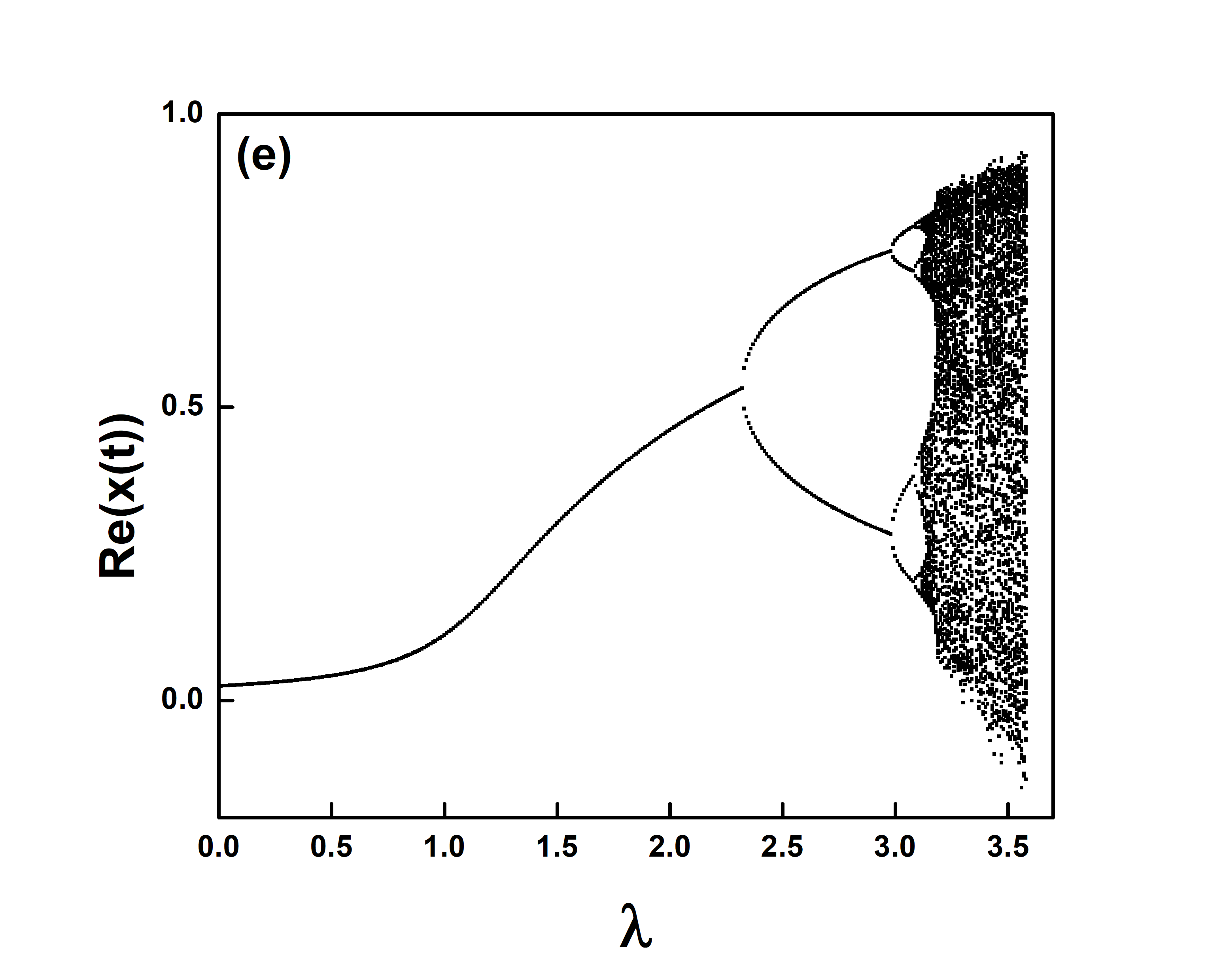}
	}
	\hfill
	\subfloat[Logistic map with $\alpha_0=0.2$, $r=0$ and $x_{0}=(0,0.01)$.]{
		\includegraphics[width=3.4in,height=2.7in]{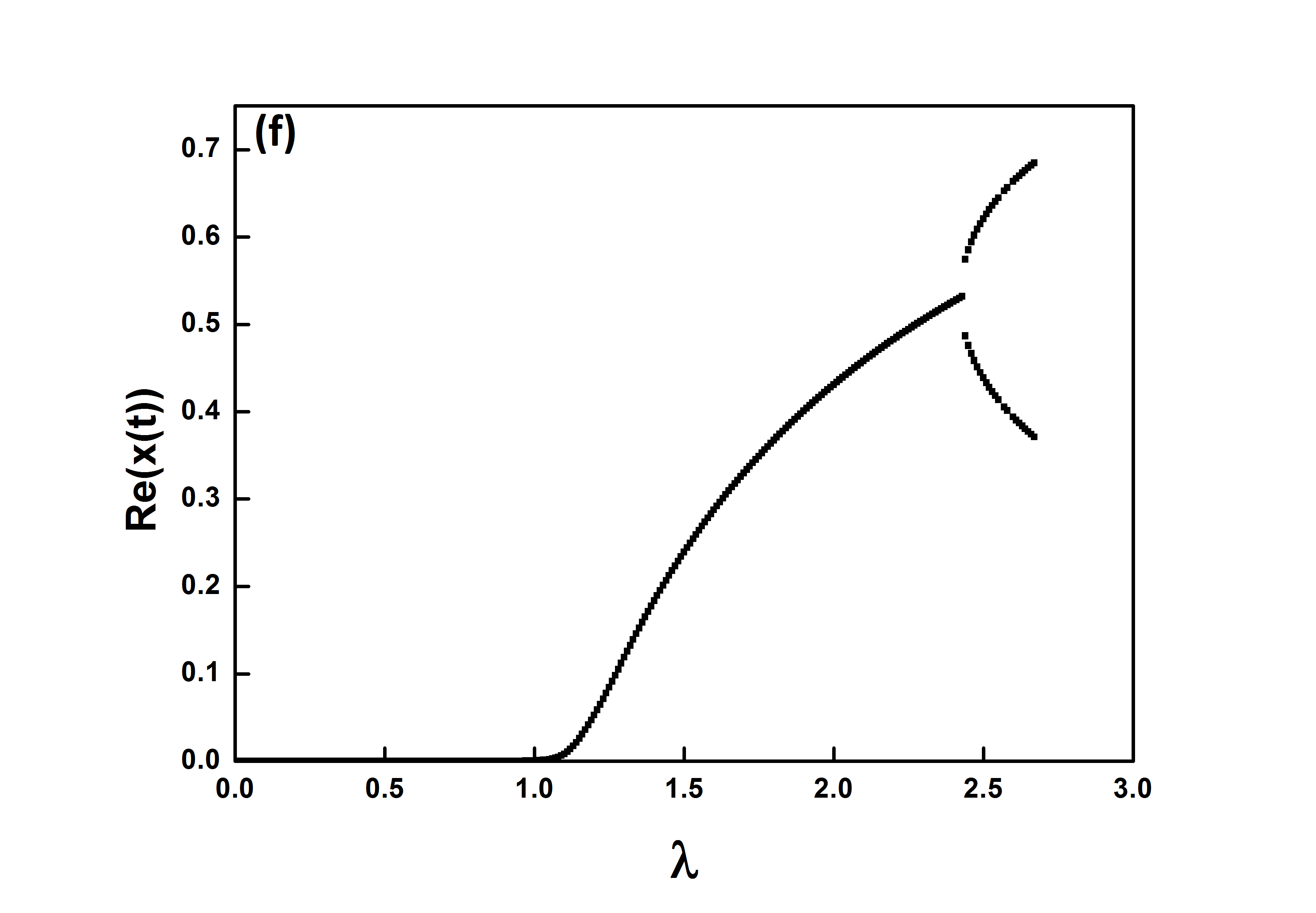}
	}
	\caption{Loss of chaos for complex initial conditions in 
		H{\'e}non, Gauss, and logistic map for real
		fractional order (b,d,f). The bifurcation diagrams for real initial 
		conditions are also shown for comparison (a,c,e).}
		\label{fige}
\end{figure*}

If the chaotic attractor is 
ergodic, different initial conditions lead to the same
attractor. In general, in integer-order $1d$ maps, we do not observe
multistability and different initial conditions
lead to the same attractor. But for maps of complex
fractional-order showing chaos, we observe multistability.
This effect may be due to extending the order of difference equations and
making $\alpha$ complex. It also could be an artifact of the fact that
the variables like $x(t)$ or $y(t)$  become complex even 
if we start with real initial conditions. To understand 
this effect, we carry out simulations for
real fractional order maps by giving
complex initial conditions.
We study the effect of complex 
initial conditions for  maps of fractional real order.
For Gauss, logistic, and H{\'e}non maps, 
we plot the bifurcation diagram for real and complex initial conditions (see Figure (\ref{fige})).
For the logistic map, we observe fixed point, period-2 and period-4 orbits. Period-2 orbits can be 
found analytically for the 
logistic map and the method is outlined in the 
appendix. We observe no multistability
in this case. We show basin of attraction for period-4 points of logistic map for $\lambda=3.29, \alpha_0=0.8, r=0.01$ in figure(\ref{figd1}).
We note that all initial conditions which do not converge to period-4 escape to infinity. The basin has an
interesting fractal structure and can be viewed as Julia set for  this system. 
Thus, there is no multistability in this system. Furthermore,
we do not observe any aperiodic attractor.
We spot a correlation between the existence of chaos 
and the initial conditions. It is a crucial revelation 
that chaos vanishes in all three continuous and differentiable 
maps (Gauss, logistic, and H{\'e}non maps) 
when the initial conditions are complex for 
real fractional order {\it{i.e.}}, for $r=0, 0<\alpha_0<1$. 
In prior studies of fractional real order maps, initial conditions were 
real constants as seen in \cite{deshpande2016chaos}.

\begin{figure*}[ht!]
	\subfloat[Circle map with $\alpha_0=0.9$, $r=0.1$ and $x(0)=0$.]{%
		\includegraphics[width=3.4in,height=2.7in]{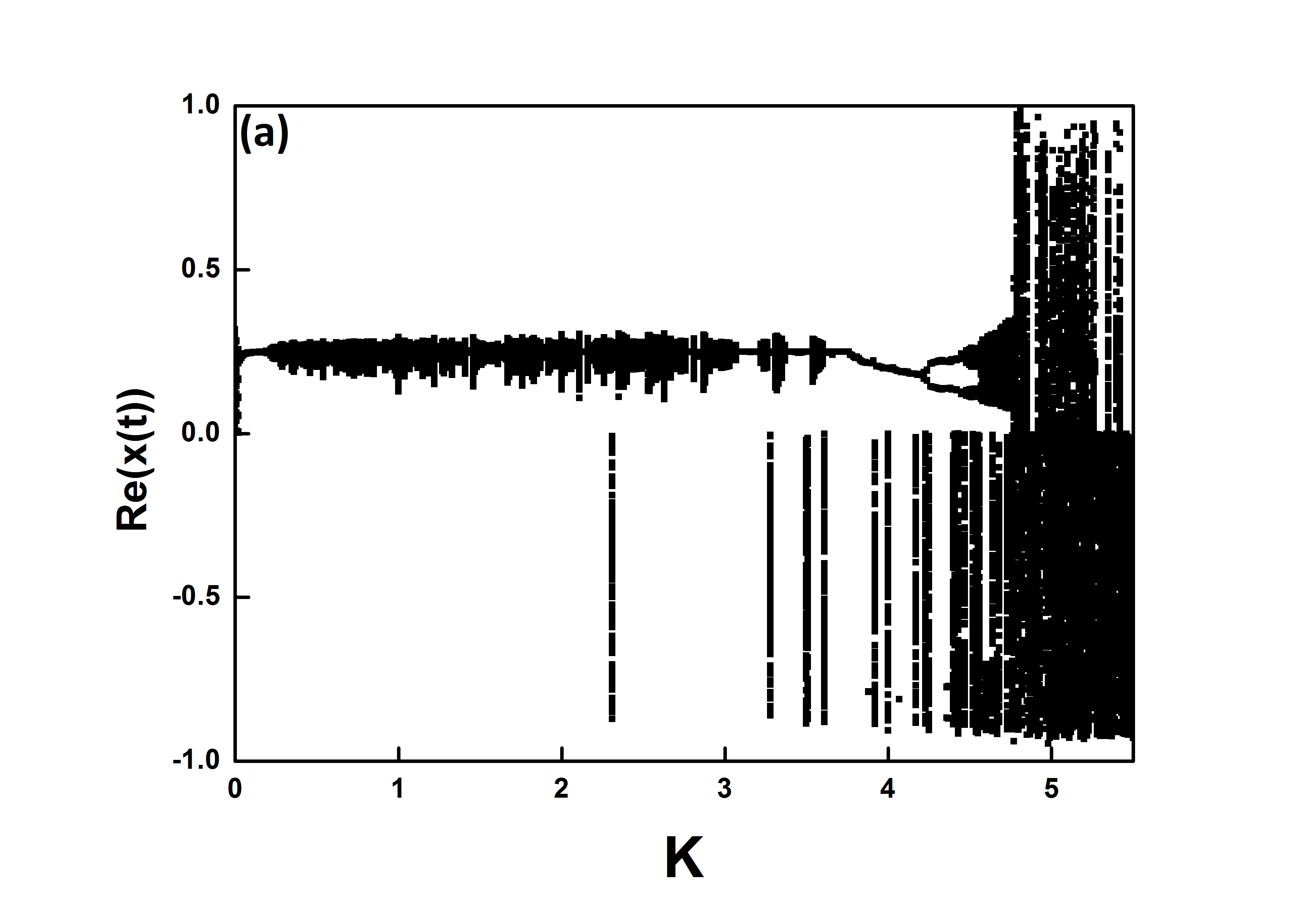}
	}
	\subfloat[Bernoulli map with $\alpha_0=0.7$, $r=0.5$ and $x(0)=0$.]{%
		\includegraphics[width=3.4in,height=2.7in]{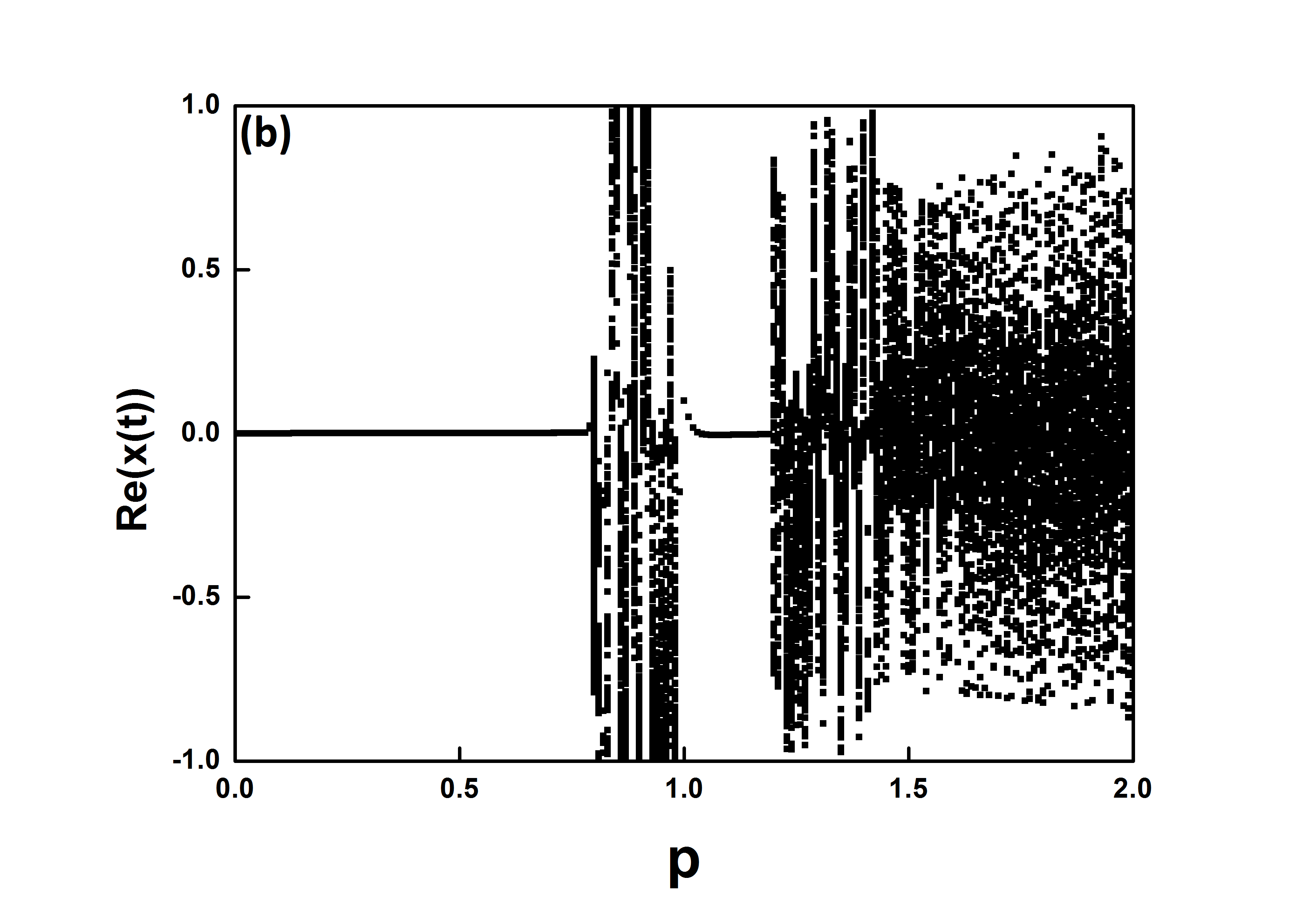}
	}\\
	\subfloat[Tent map with $\alpha_0=0.8$, $r=0.01$ and $x(0)=0$.]{%
		\includegraphics[width=3.4in,height=2.7in]{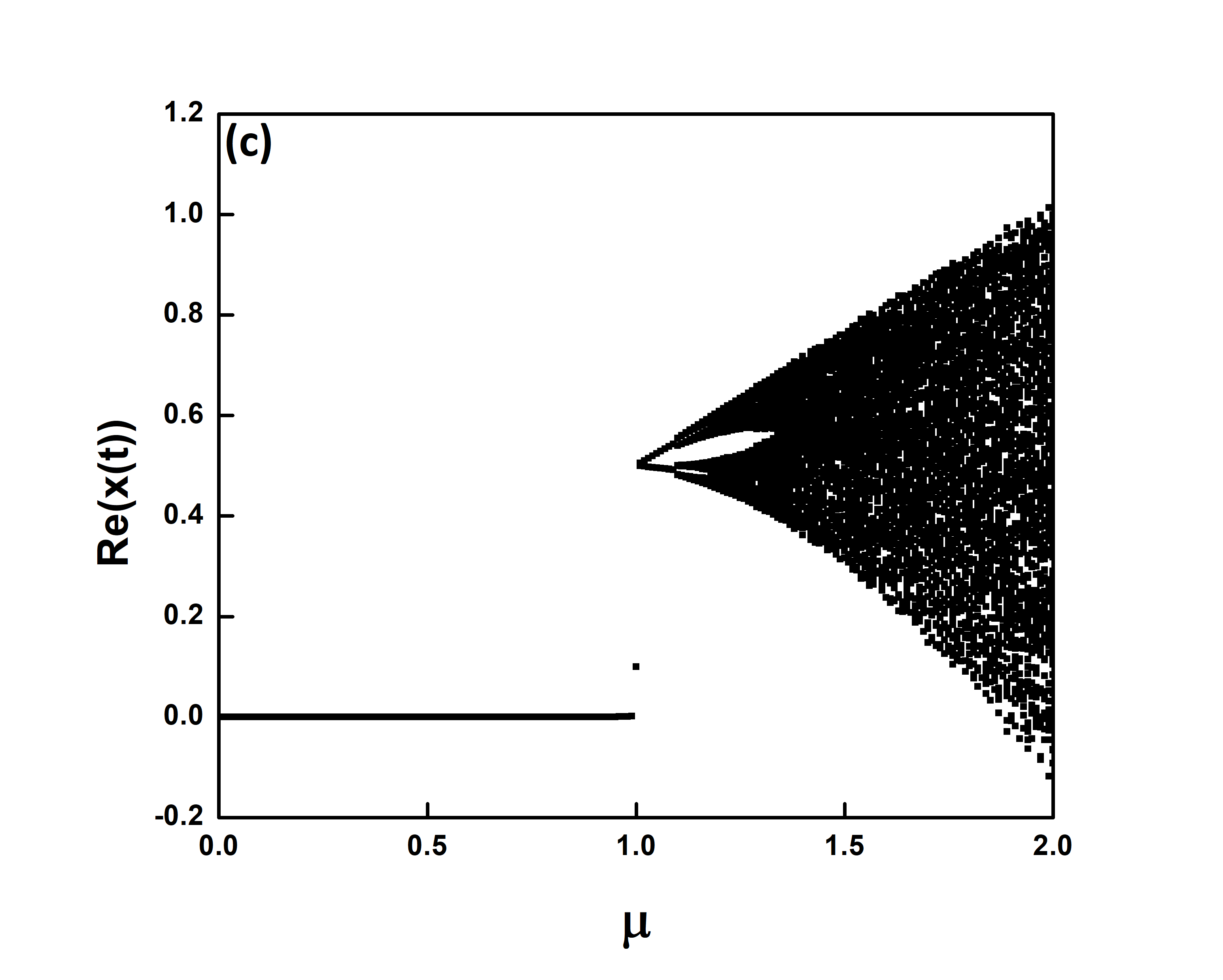}
	}
	\caption{Bifurcation diagrams for circle map, Bernoulli map, and tent map.}
	\label{figf}
\end{figure*}

We study  discontinuous maps such as the Bernoulli map and the circle map, 
as well as continuous but non-differentiable, such as the tent map. 
The bifurcation diagram (see Figure (\ref{figf})) shows that, even with the introduction of 
complex fractional-order, the chaotic attractor is not destroyed. Thus, 
we may correlate
the existence of chaos for complex fractional-order with the 
analytic nature of the maps. 
In discontinuous maps, we find chaotic attractors in all cases namely,
a) Real fractional-order and real initial conditions b) Real 
fractional-order and complex initial conditions c) Complex fractional 
order and real initial conditions and d) Complex fractional 
order and complex initial conditions. On the other hand, for H{\'e}non, 
logistic, and Gauss maps, we observe chaotic attractor only for
real fractional-order and real initial conditions.

\begin{figure*}[ht!]
	\subfloat[Lozi map with $\alpha=0.4+0.3\iota$.]{%
		\includegraphics[width=3.4in,height=2.7in]{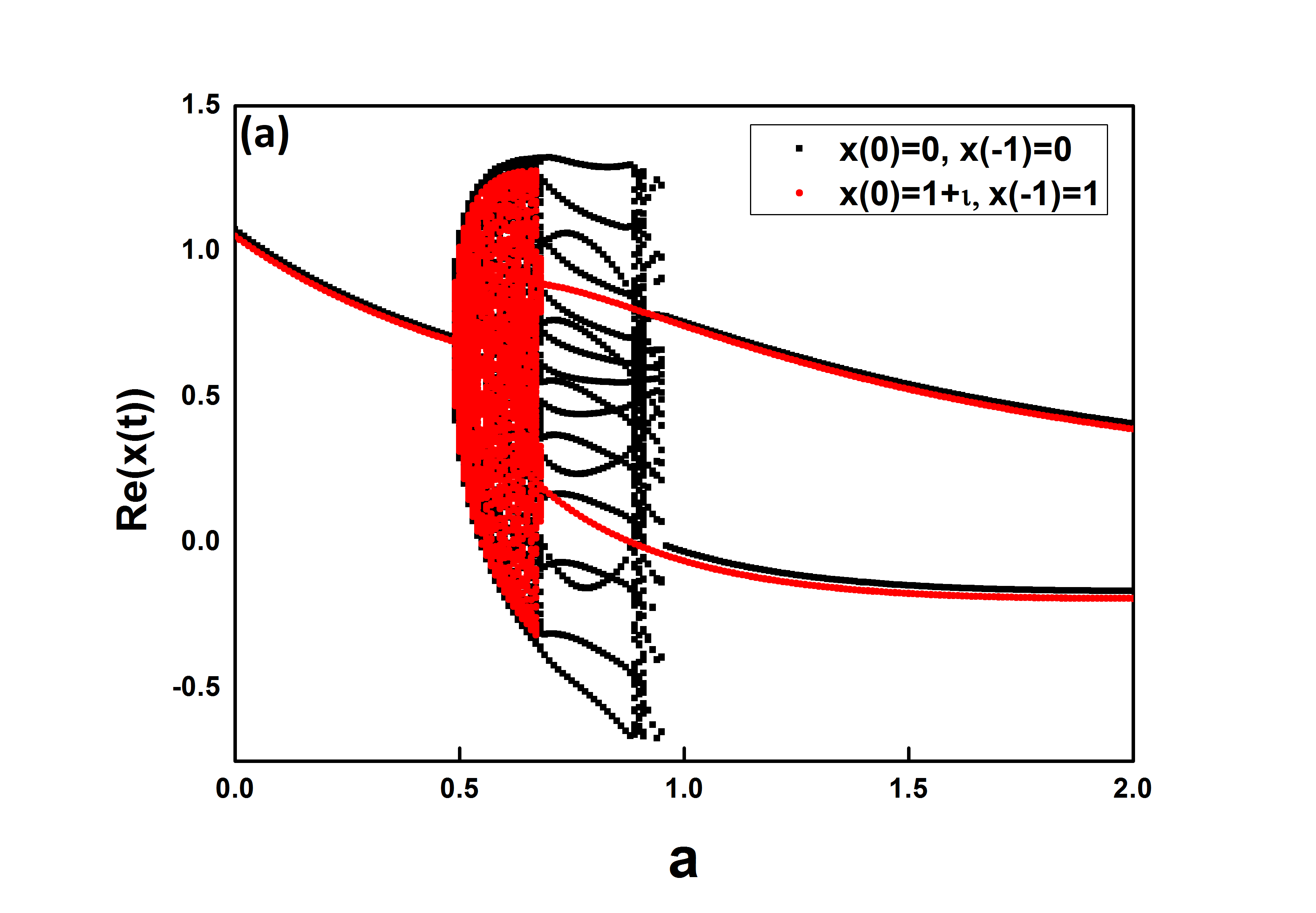}
	}
	\subfloat[ Lozi map with $\alpha=0.4+0.5\iota$.]{%
		\includegraphics[width=3.4in,height=2.7in]{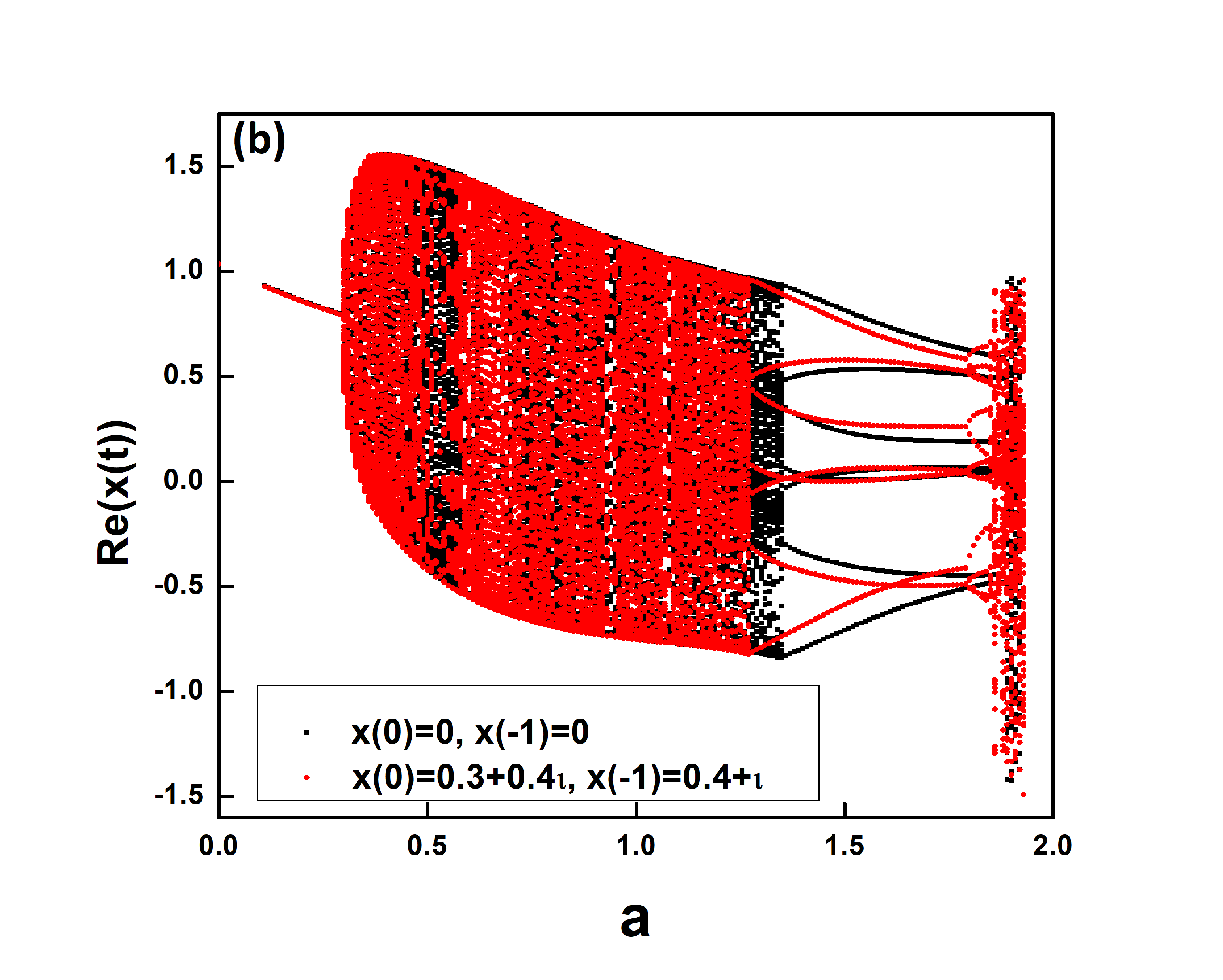}
	}\\
	\subfloat[Bernoulli map with $\alpha_0=0.8$ and $r=0.01$.]{%
		\includegraphics[width=3.4in,height=2.7in]{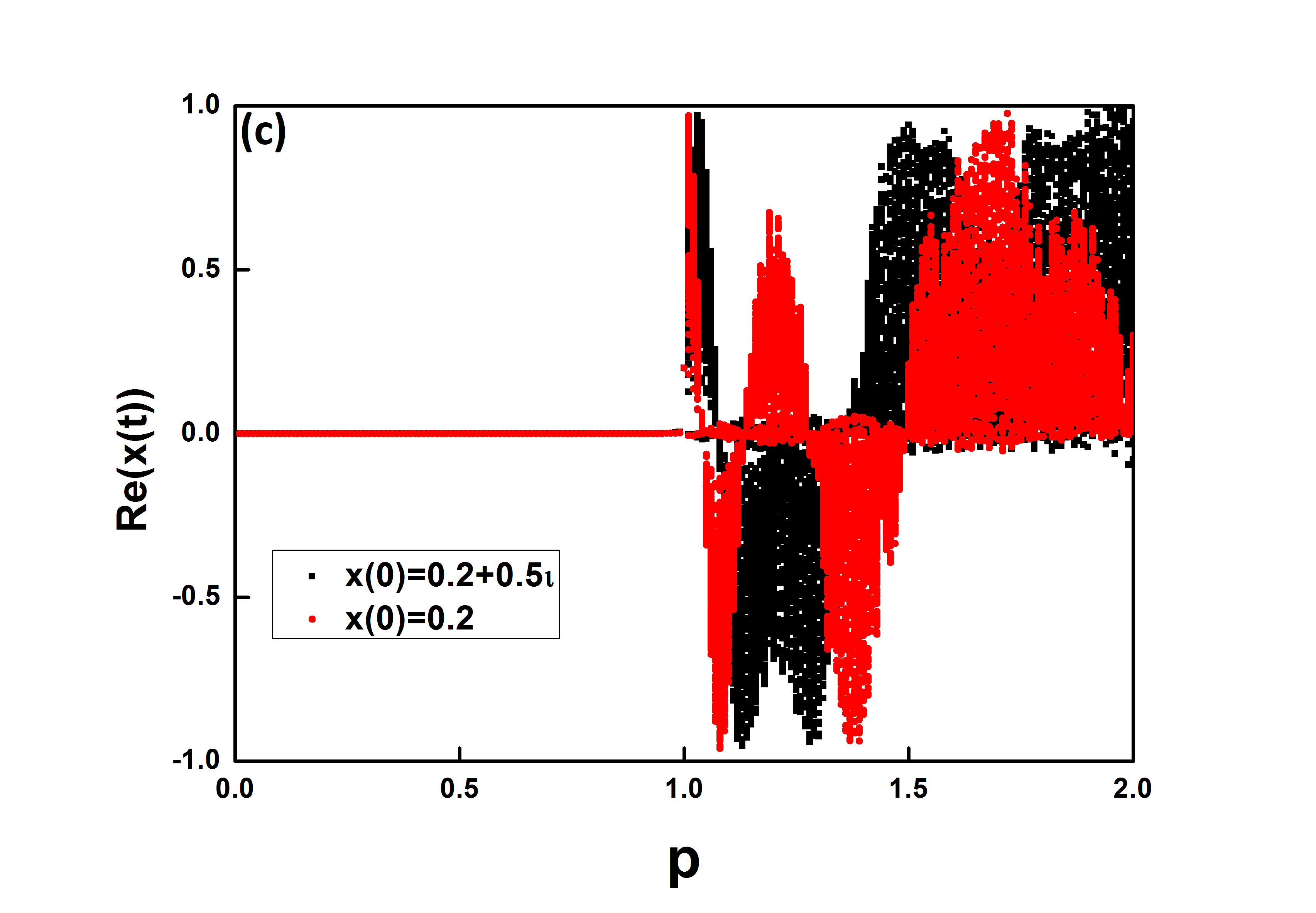}
	}
	\subfloat[Circle map with $\alpha_0=0.1$ and $r= 0.1$.]{%
		\includegraphics[width=3.4in,height=2.7in]{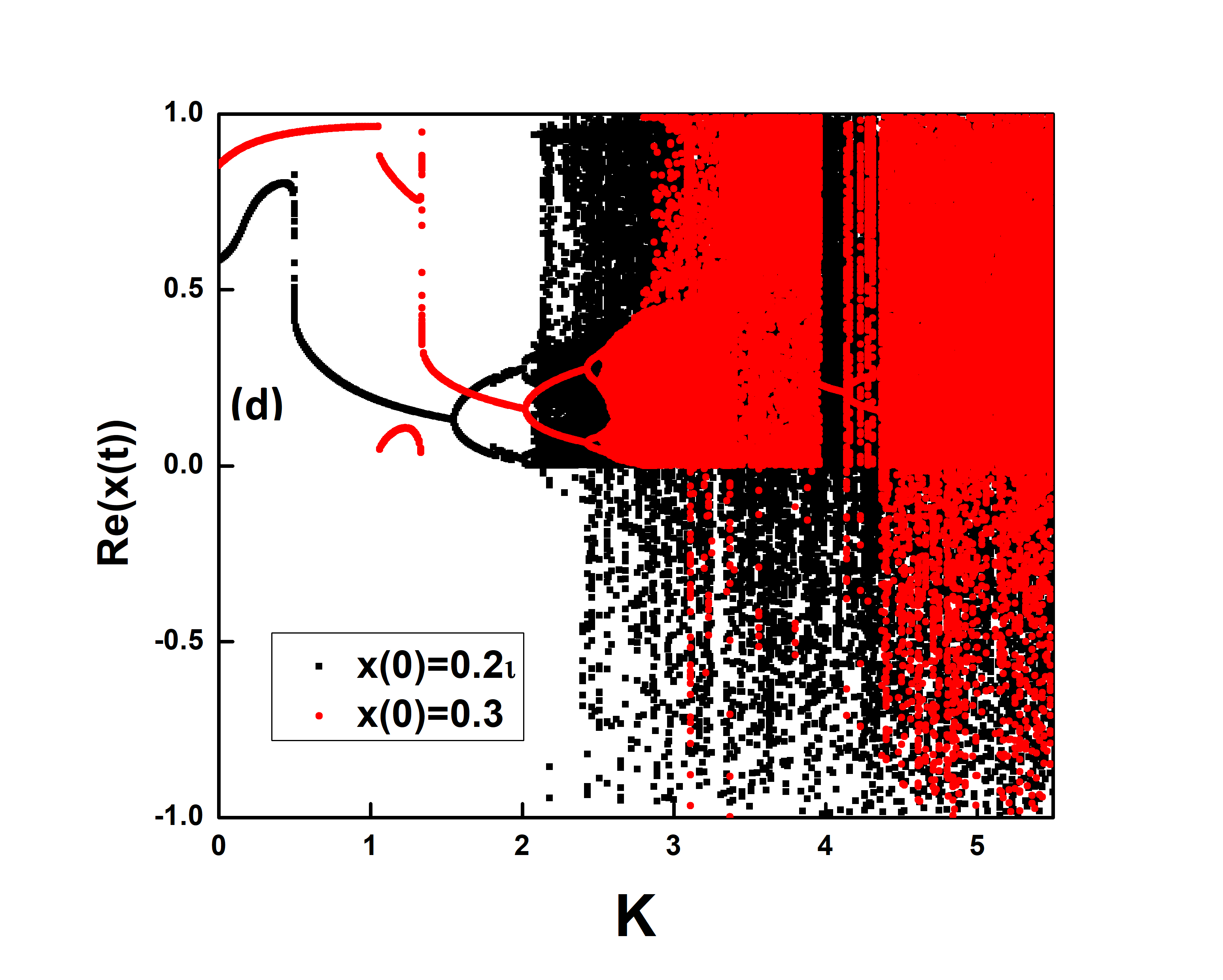}
	}\\
	\subfloat[Tent map with $\alpha_0=0.9$ and $r=0.5$.]{%
		\includegraphics[width=3.4in,height=2.7in]{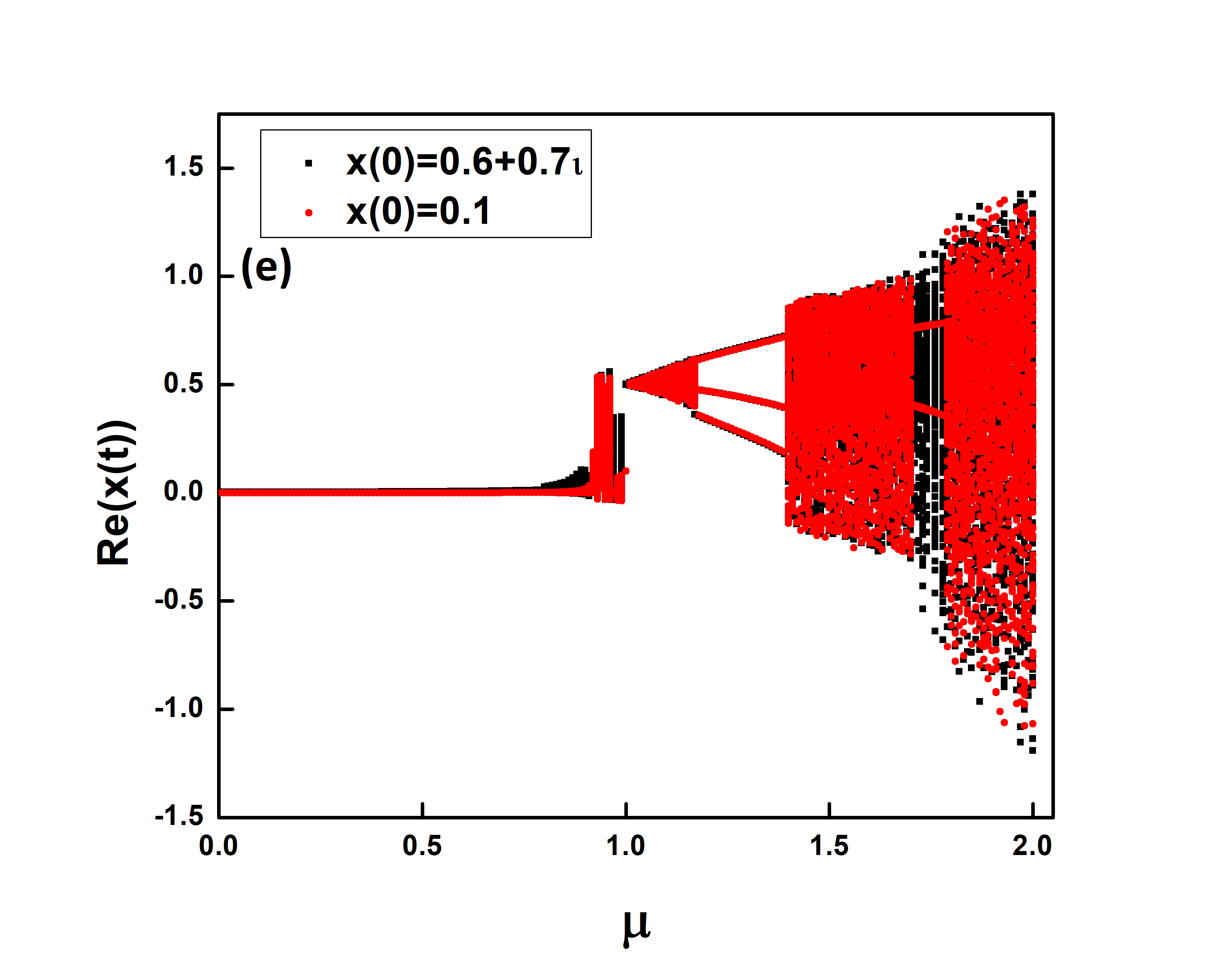}
	}
	\caption{Bifurcation diagrams for Lozi, Bernoulli, circle, and tent maps
		 	with two different initial conditions. 
		Bifurcation diagrams are different indicating clear multistability in the systems.}
	\label{figg}
\end{figure*}

We note that extremely strange bifurcations are observed
in the Lozi map 
(model L1)
of complex fractional-order. We have shown bifurcation
diagrams for $\alpha=0.4+0.3\iota$ and $\alpha=0.4+0.5\iota$
 in Figures (\ref{figg}a and \ref{figg}b) 
which clearly shows the possibility of very large periods and the 
rich bifurcation structure which
is usually not seen in integer-order systems. 
From these figures, it is clear that
 two different initial conditions lead to different bifurcation diagrams 
 for the Lozi map which 
 is a clear indication of multistability.

\begin{figure*}[ht!]
	\subfloat[Basin of attraction for Lozi map.]{%
		\includegraphics[scale=0.35]{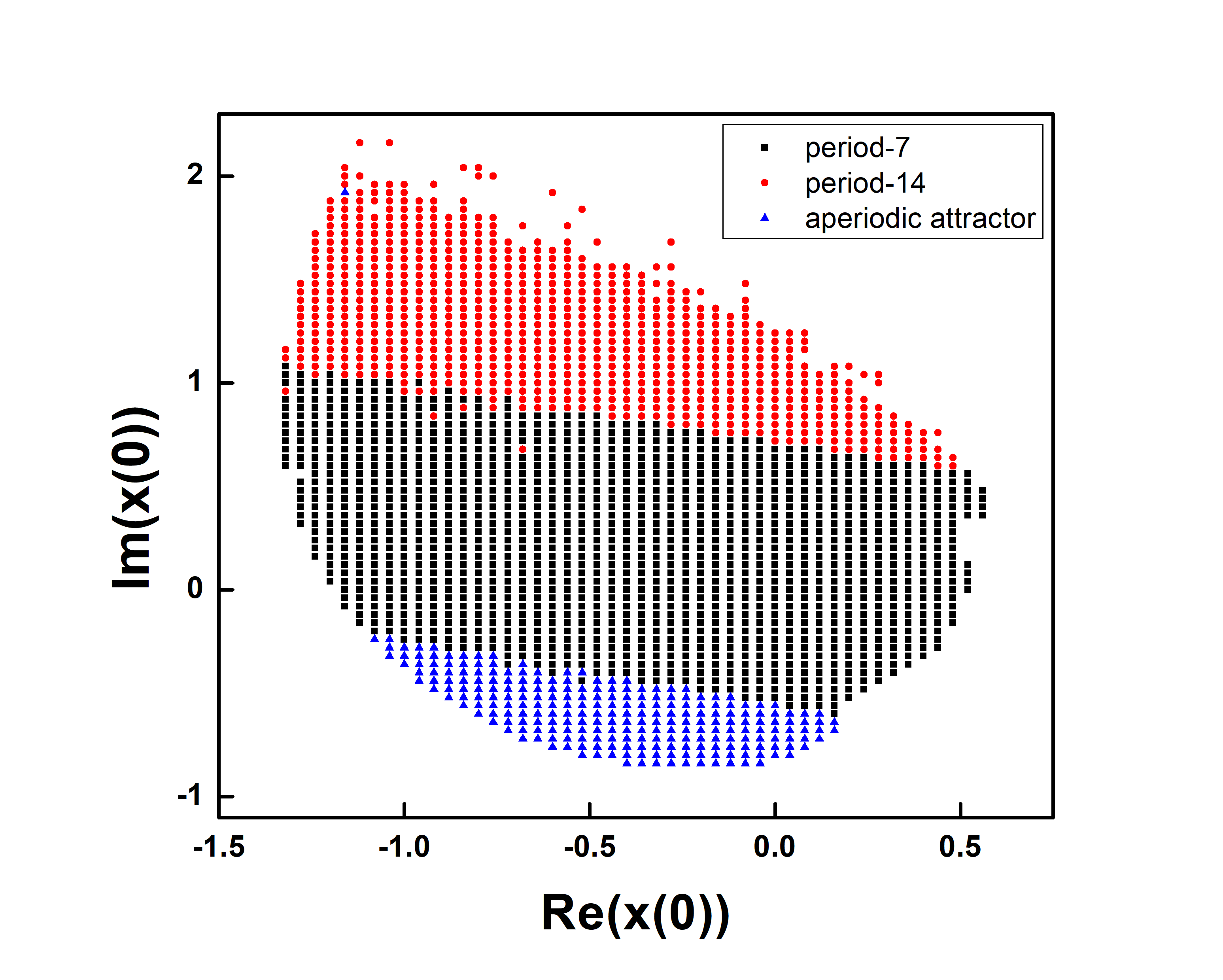}
	}
	\subfloat[Attractor of period-7 with $x(0)=-0.5+1.2\iota$.]{%
		\includegraphics[scale=0.35]{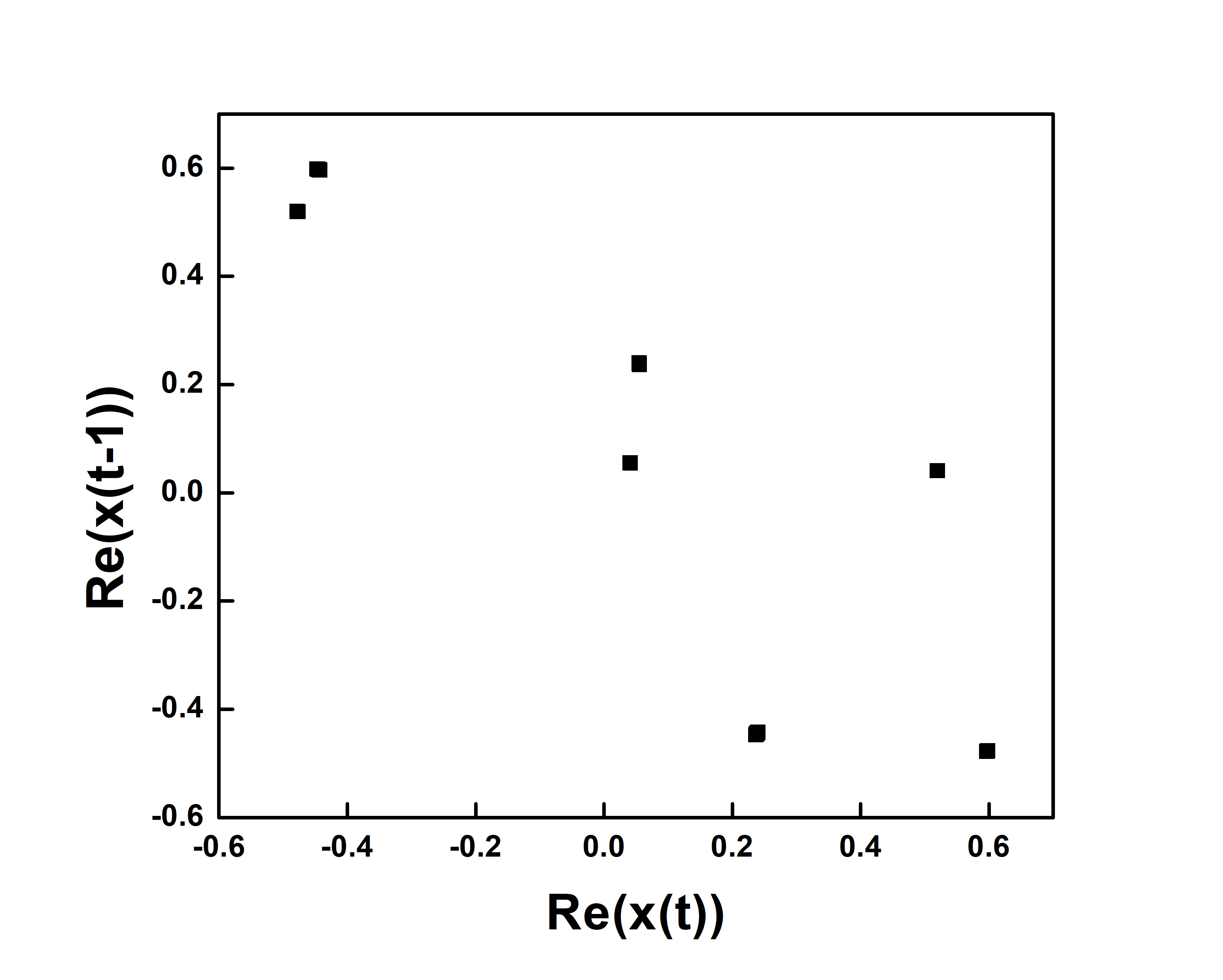}
	}\\
	\subfloat[Attractor of period-14 with $x(0)=-0.8-0.5\iota$.]{%
		\includegraphics[scale=0.35]{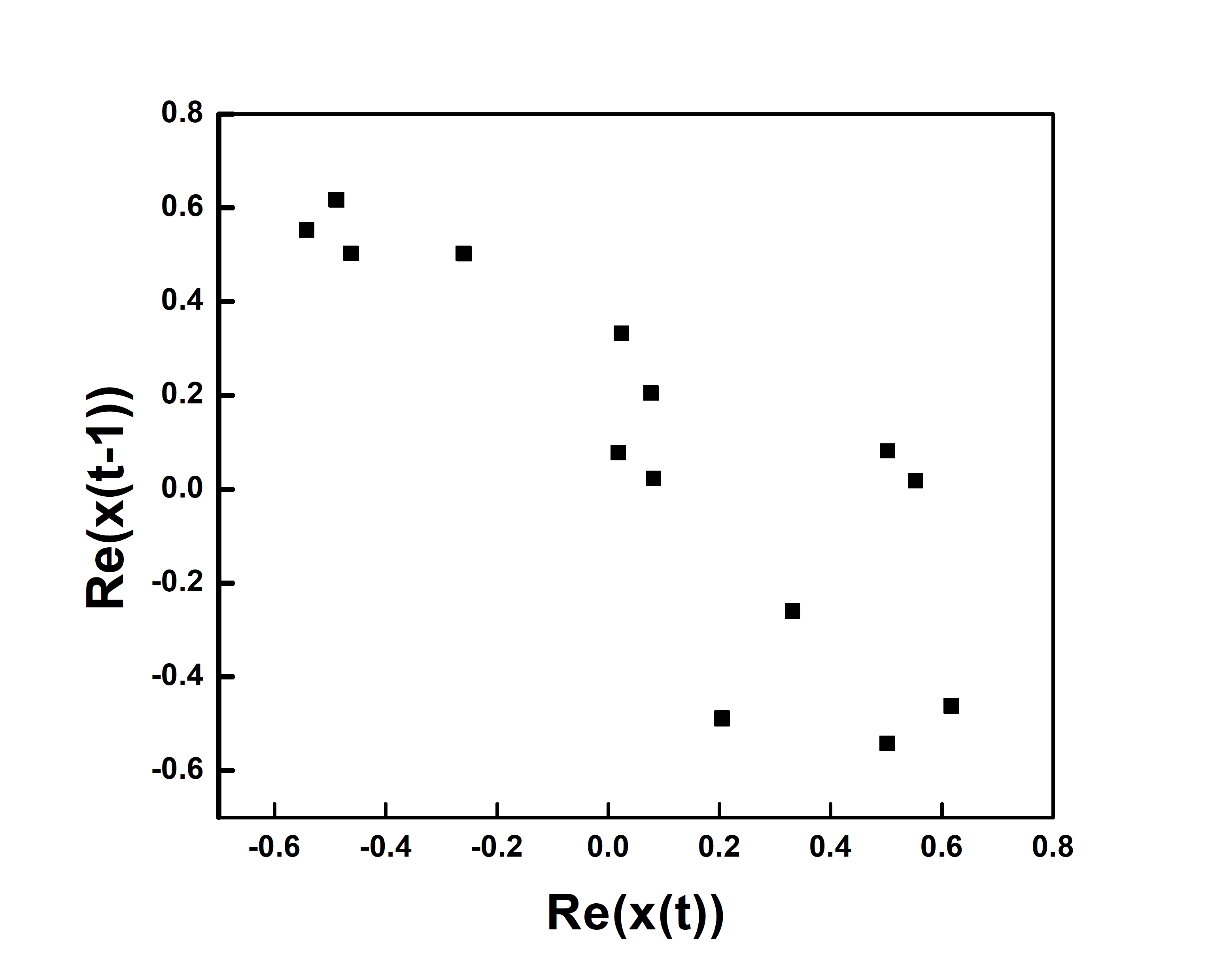}
	}
	\subfloat[Aperiodic attractor with $x(0)=0.3+0.4\iota$.]{%
		\includegraphics[scale=0.35]{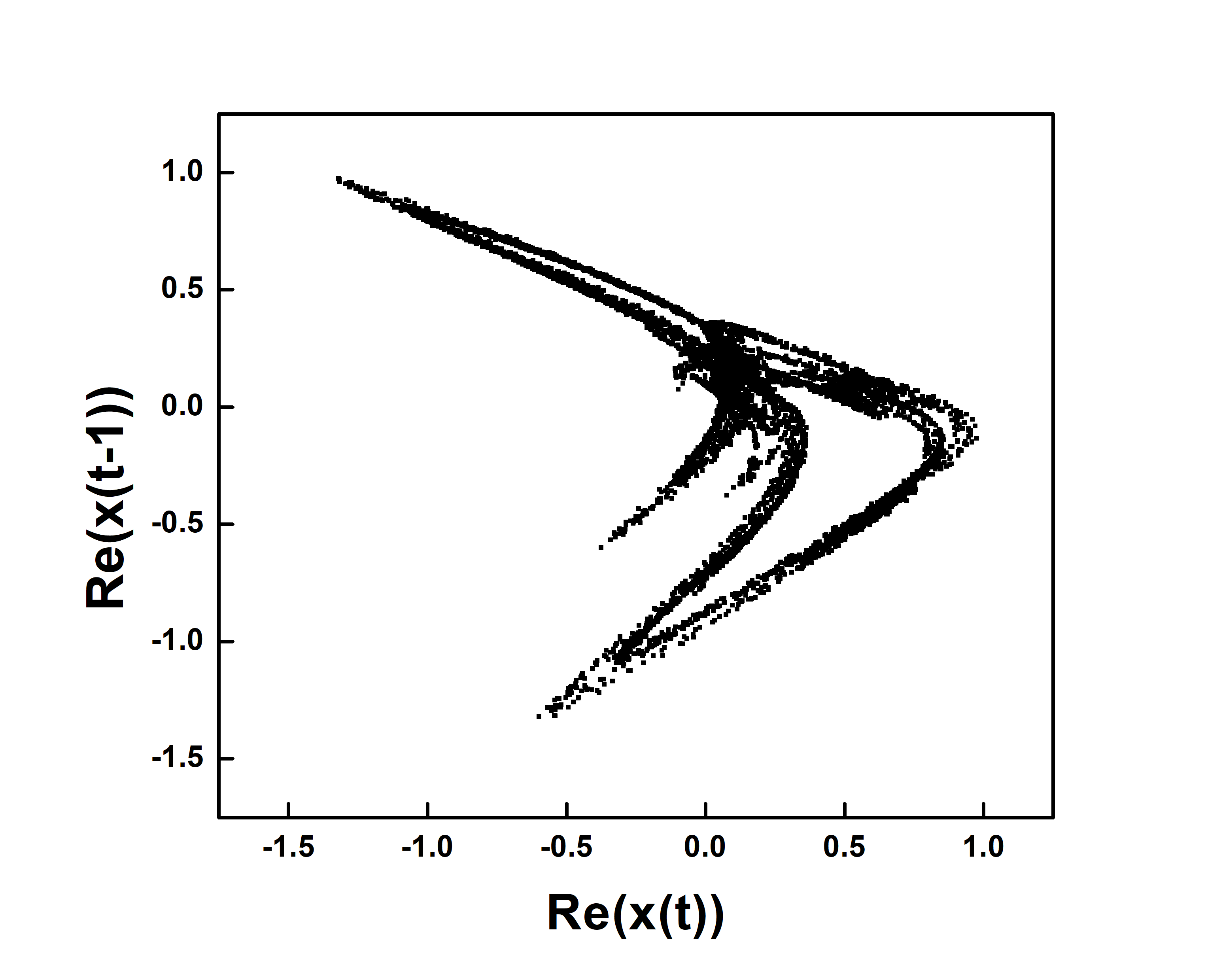}
	}
	\caption{Lozi map showing multiple attractors for varying combinations of initial conditions.}
	\label{figf1}
\end{figure*}

For Lozi map, for $x(-1)=0.4+\iota$, $\alpha_0=0.4+0.5\iota, 
a=1.8, b=0.05$ three different attractors are realized for different
initial values of $x(0)$. We observe period-7, period-14, and 
aperiodic attractor. These basins and attractors 
are shown in figure(\ref{figf1}).

We have also shown the bifurcation diagram with two different
 initial conditions for Bernoulli, circle, 
and tent map in Figures (\ref{figg}c, \ref{figg}d, and 
\ref{figg}e). It is clear that the bifurcation 
diagram changes. Thus, different attractors are reached 
with different initial conditions. Thus, these maps 
clearly show multistability. We studied basins of attraction
of different attractors in circle map. For $a=4.4$, we observe
both period-2 and aperiodic attractor. However, basins of
attractors are intermingled and initial conditions leading 
to periodic attractor are randomly distributed and do not 
have a smooth or fractal structure.

If the map is continuous and differentiable, we have
 another observation. For real $\alpha$, {\it{i.e.}} for
$r=0$ if there is a chaotic attractor for real order, the 
system goes to infinity for complex order even for $r=0.001$ 
and the bifurcation diagram shows a gap at these parameter values.

Apart from these maps, we also studied the cubic map in $1d$ 
\cite{rogers1983chaos} given as:
$$ x_{n+1}=ax_{n}^{3}+(1-a)x_{n}.$$
and Duffing map in $2d$ defines \cite{ouannas2020discrete}:
$$x_{n+1}=y_{n},$$
$$y_{n+1}=-bx_{n}+ay_{n}-y_{n}^{3}.$$
The right-hand side of both these functions
is an analytic function even if extended to a complex domain.
We study the bifurcation diagrams for
the real value of $\alpha$ with 
complex initial conditions and also maps of complex fractional order
with a nonzero imaginary part. In either case, we do not see any chaos
over a range of parameters studied. Thus, we have five cases 
(Gauss map, logistic map, H{\'e}non map, Duffing map, and cubic map) where
we have identical observations, namely, with complex initial and real fractional 
order or with complex fractional order, we do not observe any chaos
over the range of parameters studied.

\section{Results and Conclusions}
We have studied $1d$ and $2d$ maps of complex fractional-order. 
In $2d$, we studied Lozi and H{\'e}non maps of complex fractional-order. There are two possible generalizations
and we have studied one of them in detail.
We find that the chaos disappears completely when we
introduce a small imaginary part to fractional-order $\alpha$ for
the H{\'e}non map. For the Lozi map, the chaos does not disappear completely and
is indeed seen for some parameter values even for complex $\alpha$.
We also note that the system has memory. H{\'e}non
map does not show 
pronounced multistability for complex order
as seen in the original map and the bifurcation
diagrams are almost independent of the initial conditions as
long as the initial condition is stable. On the other hand, the Lozi
map shows multistability and the asymptotic attractor is dependent on the initial conditions.

Thus, there are qualitative
differences in the dynamical behavior of 
complex fractional-order H{\'e}non and 
Lozi  maps. We believe that these results are 
interesting and further studies are essential to
understand the qualitative differences in behaviors of different
maps of fractional-order.

We try to study the possibility of chaos in maps of complex fractional order. 
In $1d$, we studied logistic, tent, Gauss, circle, and Bernoulli maps.
These studies along with studies in H{\'e}non and
Lozi  maps suggest that maps that are continuous and 
differentiable lose chaos for real fractional order when
the initial conditions are complex and they do not show 
chaos at all for any initial condition when the order
is complex. Whereas, for maps that are discontinuous or 
non-differentiable we observe chaos for some initial conditions
and parameter values.
Multistability is seen for Lozi, Bernoulli, circle, 
and tent maps of complex fractional-order. Thus, multistability 
can be observed even $1d$ fractional difference equations 
of complex order if the underlying map is
not analytic.

Often, non-linearity is necessary condition for chaos. 
In all the examples of complex order fractional
difference equations we studied, it is striking 
that we required non-analyticity as well.

\section{Acknowledgement}
PMG and DDJ thank DST-SERB for financial assistance 
(Ref. CRG/2020/003993). SB acknowledges the University 
of Hyderabad for Institute of Eminence-Professional 
Development Fund (IoE-PDF) by MHRD (F11/9/2019-U3(A)).

Author Credits:  DDJ, PMG and SB contributed equally to 
the main paper. SB: Appendix A.


%
%

%


\appendix

\section{Period-2 points in fractional order maps}

Following the method outlined in \cite{edelman2014fractional}, we compute the period-two points for logistic map. 

Consider,
\begin{equation}
	x(t)=x(0)+\frac{1}{\Gamma(\alpha)}\sum_{j=1}^{t}\frac{\Gamma(t-j+\alpha)}{\Gamma(t-j+1)}[f(x(j-1))-x(j-1)], \label{eq6}
\end{equation}
where, $0<\alpha<1$.
If there exist the points $p$ and $q$ such that \[\lim_{t\to\infty}x(2t)=p\]  and \[\lim_{t\to\infty}x(2t+1)=q,\] $p \ne q$ then we say that the system (\ref{eq6}) has period-2 limit cycle.
We have,
\begin{equation*}
	x(2t)=x(0)+\frac{1}{\Gamma(\alpha)}\sum_{j=1}^{2t}\frac{\Gamma(2t-j+\alpha)}{\Gamma(2t-j+1)}[f(x(j-1))-x(j-1)]
\end{equation*}
and 
\begin{eqnarray*}
	x(2t+1)&=&x(0)+\frac{1}{\Gamma(\alpha)}\sum_{j=1}^{2t+1}\frac{\Gamma(2t+1-j+\alpha)}{\Gamma(2t-j+2)}\\
	&& \times[f(x(j-1))-x(j-1)].
\end{eqnarray*}
Therefore,
\begin{eqnarray}
	x(2t)&=&x(0)+\sum_{k=0}^{t-1}\frac{\Gamma(2t-2k-1+\alpha)}{\Gamma(2t-2k)\Gamma(\alpha)}[f(x(2k))-x(2k)]\nonumber \\ 
	&&+\sum_{k=0}^{t-1}\frac{\Gamma(2t-2k-2+\alpha)}{\Gamma(2t-2k-1)\Gamma(\alpha)} \nonumber \\
	&&\times[f(x(2k+1))-x(2k+1)]  \label{eq7} 
\end{eqnarray}
and
\begin{eqnarray}
	x(2t+1)&=&x(0)+\sum_{k=0}^{t}\frac{\Gamma(2t-2k+\alpha)}{\Gamma(2t-2k+1)\Gamma(\alpha)}[f(x(2k))-x(2k)]\nonumber \\
	&&+\sum_{k=0}^{t-1}\frac{\Gamma(2t-2k-1+\alpha)}{\Gamma(2t-2k)\Gamma(\alpha)} \nonumber \\
	&& \times[f(x(2k+1))-x(2k+1)].  \label{eq8}
\end{eqnarray}
Taking $\lim_{t\to\infty}$ and subtracting (\ref{eq8}) from (\ref{eq7}), we get
\begin{eqnarray*}
	p-q&=&(f(p)-p)\lim_{t\to\infty}\left(\sum_{k=0}^{t-1}\frac{\Gamma(2t-2k-1+\alpha)}{\Gamma(2t-2k)\Gamma(\alpha)}\right. \\
	&&\left.-\sum_{k=0}^{t}\frac{\Gamma(2t-2k+\alpha)}{\Gamma(2t-2k+1)\Gamma(\alpha)}\right)\\
	&&+(f(q)-q) \lim_{t\to\infty}\left(\sum_{k=0}^{t-1}\frac{\Gamma(2t-2k-2+\alpha)}{\Gamma(2t-2k-1)\Gamma(\alpha)}\right.\\
&&\left.-\sum_{k=0}^{t-1}\frac{\Gamma(2t-2k-1+\alpha)}{\Gamma(2t-2k)\Gamma(\alpha)}\right)
\end{eqnarray*}
The term $\sum_{k=0}^{t-1}\frac{\Gamma(2t-2k-2+\alpha)}{\Gamma(2t-2k-1)\Gamma(\alpha)}$ can be replaced with $\sum_{k=0}^{t}\frac{\Gamma(2t-2k+\alpha)}{\Gamma(2t-2k+1)\Gamma(\alpha)}$ if we put $k+1$ as $k$.
Furthermore, 
\begin{eqnarray*}
&&\lim_{t\to\infty}\left(\sum_{k=0}^{t-1}\frac{\Gamma(2t-2k-1+\alpha)}{\Gamma(2t-2k)\Gamma(\alpha)}-\sum_{k=0}^{t}\frac{\Gamma(2t-2k+\alpha)}{\Gamma(2t-2k+1)\Gamma(\alpha)}\right)\\
&=& \lim_{t\to\infty}\sum_{k=0}^{2t}(-1)^{k+1}\frac{\Gamma(2t-k+\alpha)}{\Gamma(2t-k+1)\Gamma(\alpha)}\\
&=& -\sum_{k=0}^{\infty}
\begin{pmatrix}
	-\alpha \\
	k
\end{pmatrix} \nonumber\\
&=& -2^{-\alpha}.
\end{eqnarray*}
Therefore,
\begin{eqnarray}
	p-q=-[(f(p)-p)-(f(q)-q)]2^{-\alpha}. \label{eq9}
\end{eqnarray}
Now adding (\ref{eq7}) and (\ref{eq8}) and taking $\lim_{t\to\infty}$ we get
\begin{eqnarray}
	p+q&=&2x(0)+2[(f(p)-p)+(f(q)-q)] \nonumber \\
	&&\times[\lim_{t\to\infty}\sum_{k=0}^{2t}\frac{\Gamma(2t-k+\alpha)}{\Gamma(2t-k+1)\Gamma(\alpha)}]. \nonumber
\end{eqnarray}
Since the limit in the above equation tends to infinity, and all other terms are finite, we must have
\begin{equation}
	p+q=f(p)+f(q).  \label{eq10}
\end{equation}
Solving equations (\ref{eq9}) and (\ref{eq10}), we get 
\begin{eqnarray}
	p&=&\left(\frac{1}{2}-\frac{2^{-\alpha-1}}{1-2^{-\alpha}}\right)f(p)+\left(\frac{1}{2}+\frac{2^{-\alpha-1}}{1-2^{-\alpha}}\right)f(q)\nonumber \\
	q&=&\left(\frac{1}{2}+\frac{2^{-\alpha-1}}{1-2^{-\alpha}}\right)f(p)+\left(\frac{1}{2}-\frac{2^{-\alpha-1}}{1-2^{-\alpha}}\right)f(q).\nonumber \\
	&& \label{eq11}
\end{eqnarray}
Let us consider the case of fractional order logistic map with period-2.
Here, $f(x)=\lambda x(1-x)$, $0<\alpha<1$, $\lambda \in [0,4]$. Solving equation for (\ref{eq11}), we get the equilibrium points :
$$(p,q)=(0,0),$$
$$(p,q)=(\frac{\lambda-1}{\lambda},\frac{\lambda-1}{\lambda})$$
and 
\begin{eqnarray}
	(p,q)&=&\left(\frac{(2^{\alpha}-1)+\lambda \pm \sqrt{(\lambda-(1-2^{\alpha}))(\lambda-(1+2^{\alpha}))}}{2\lambda},\right.\nonumber \\
	&&\left. \frac{(2^{\alpha}-1)+\lambda \mp \sqrt{(\lambda-(1-2^{\alpha}))(\lambda-(1+2^{\alpha}))}}{2\lambda}\right). \nonumber
\end{eqnarray}

For $\alpha=1$
\begin{eqnarray}
	p&=&\frac{1+\lambda \pm \sqrt{(\lambda+1)(\lambda-3)}}{2\lambda}\nonumber \\
	q&=&\frac{1+\lambda \mp \sqrt{(\lambda+1)(\lambda-3)}}{2\lambda}.\nonumber \\
\end{eqnarray}
which matches with integer order logistic map \cite{lakshmanan2012nonlinear}.

Thus, for real values of roots there exists 2-limit cycle in the fractional order logistic map if $\lambda$ satisfies $(\lambda-(1-2^{\alpha}))(\lambda-(1+2^{\alpha}))>0,$ $0<\alpha<1$.
That is if $\lambda>1+2^{\alpha}$ or $\lambda<1-2^{\alpha}$. Since $\lambda\in[0,4],$ we must have $\lambda\in[1+2^{\alpha},4],$ for 2-cycle.

\bibliography{complex}
\bibliographystyle{ieeetr}

\end{document}